\documentclass[12pt,letterpaper]{article}

\usepackage[round]{natbib}
\usepackage{graphicx}

\usepackage{url}
\usepackage{enumerate}
\usepackage{bm}
\usepackage{color}
\usepackage{palatino}
\usepackage{csvsimple}
\usepackage{subfigure}
\usepackage{fullpage}
\usepackage{amsmath,amsfonts,amssymb,amsthm}
\usepackage[labelfont=bf,labelsep=period,font=small]{caption} 
\usepackage{todonotes}

\usepackage{xr}
\newcommand{\PP}{\mathbb{P}}
\newcommand{\EE}{\mathbb{E}}
\newcommand{\RR}{\mathbb{R}}
\newcommand{\ZZ}{\mathbb{Z}}
\newcommand{\Multinom}{\operatorname{Multinom}}
\newcommand{\Dirichlet}{\operatorname{Dirichlet}}

\newcommand{\arxiv}[1]{#1}
\newcommand{\notarxiv}[1]{}
\newcommand{\eat}[1]{}

\newcommand{\EM}[1]{\textit{\colorbox{magenta}{Erick:} #1}}
\newcommand{\CW}[1]{\textit{\textcolor{white}{\colorbox{blue}{Chris:}} #1}}

\newcommand{\dummyinput}[1]{}

\dummyinput{
	\input{dataset_table.csv}
	\input{topo_gr.csv}
	\input{difficulty_table.csv}
	\input{ds1_convergence.csv}
	\input{multi_table.csv}
	\input{mean_cover_all.csv}
	\input{dataset_beiko.csv}
	\input{legacy_treebase.csv}
}

\newcommand{\TABdatahohna}{\
\begin{table}
\centering
\small
\caption{The data sets used in this study, DS1-11 (eukaryote) and VL1-6 (bacterial/archaeal).
N~= number of species; Cols = number of nucleotides; Est error = Estimated maximum standard error of split frequencies in golden runs (in \%); rDNA = ribosomal DNA; rRNA = ribosomal RNA; mtDNA = mitochondial DNA; COII = cytochrome oxidase subunit II GARTFase = phosphoribosylglycinamide formyltransferase 2.
}
	\csvreader[tabular=lrrllr,
			head to column names = false,
			table head=\hline Data & N & Cols & Type of data & Study & Est error \\\hline,
			late after line=\\,
			late after last line=\\\hline]%
		{"dataset_table.csv"}%
		{}%
		{\csvcoli & \csvcolii & \csvcoliii & \csvcoliv & \citet{\csvcolv} & \csvcolvi}%
\label{TABdatahohna}
\label{TABdatabeiko}
\end{table}
}

\newcommand{\TABgoldenconvergence}{\
\begin{table}
\centering
\small
\caption{Convergence diagnostics for the golden runs on eukaryotic datasets as reported by the MrBayes \texttt{sumt} and \texttt{sump} tools.
	We report the mean log likelihood ($\mu$LL), standard error of log likelihoods (Est LL error), maximum standard deviation of split frequencies (maxSDSF), maximum topological Gelman-Rubin potential scale reduction factor for splits (maxPSRF), and the minimum estimated sample size for the treelength parameter (ESS).
}
	\csvreader[tabular=llllll,
			head to column names = false,
			table head=\hline Data & $\mu$LL & Est LL error & maxSDSF & maxPSRF & ESS \\\hline,
			late after line=\\,
			late after last line=\\\hline]%
		{"golden_convergence.csv"}%
		{}%
		{\csvcoli & \csvcolii & \csvcoliii & \csvcoliv & \csvcolv & \csvcolvi}%
\label{TABgoldenconvergence}
\end{table}
}

\newcommand{\TABmulti}{\
\begin{table}
\centering
\caption{Sets of identical sequences and multifurcations for each dataset with at least one multifurcation in the majority rule consensus tree.
Multifurcations increase the size of the true credible set by the factor noted.
}
	\csvreader[tabular=lllr,
			head to column names = false,
			table head=\hline Data & Identical Sequences & Multifurcations & Inflation Factor \\\hline,
			late after line=\\,
			late after last line=\\\hline]%
		{"multi_table.csv"}%
		{}%
		{\csvcoli & \csvcolii & \csvcoliii & \csvcoliv}%
\label{TABmulti}
\end{table}
}

\newcommand{\TABdifficulty}{\
\begin{table}
\centering
\caption{A comparison of dataset difficulty and posterior shape parameters.
The first three columns show the mean number of iterations required to reach ASDSF less than 0.01 ($\mu$Iter) using the MrBayes default parameters (4 runs, 2 chains) as well as the resulting mean maximum split frequency error ($\mu$MaxErr) and mean split frequency RMSD ($\mu$RMSD) as compared to the golden runs.
From the golden runs, we considered properties of the \emph{top trees}---the at most 4096 highest probability trees from the 95\% credible set.
We inferred the SPR radius (Radius) which we define as the maximum SPR distance from each tree in the 95\% credible set to the topology with highest posterior probability (radius of the top trees in brackets), the size of the 95\% credible set (95CI) , the cumulative posterior probability of the top trees (Cred), and the presence of peaks.
Note that our credible set clearly underestimates the true credible set size when it exceeds the number of samples (e.g. DS9 and DS11).
``-U'' data sets include only one member from each set of identical sequences.
Note that each golden run contained 750,000 samples.
}
	\csvreader[tabular=lrrrrrrc,
			head to column names = false,
			table head=\hline Data & $\mu$Iter& $\mu$MaxErr & $\mu$RMSD & Radius & 95CI & Cred & Peaks \\\hline,
			late after line=\\,
			late after last line=\\\hline]%
		{"difficulty_table.csv"}%
		{}%
		{\csvcoli & \csvcolii & \csvcoliii & \csvcoliv & \csvcolv & \csvcolvi & \csvcolvii & \csvcolviii}%

\label{TABdifficulty}
\end{table}
}

\newcommand{\TABdsoneconvergence}{\
\begin{table}
\centering
\caption{A detailed look at performance on dataset DS1 with and without MCMCMC using varying number of runs.
The number of replicates (out of 10) with a given number of chains (Ch) are shown which found both peaks (Peak), converged to an RMSD at most 0.02 (Conv), or exceeded the iteration limit (Lim).
The mean number of iterations, running time (iterations*chains*runs), maximum split frequency error (MaxErr), and RMSD are also shown.
}
	\csvreader[tabular=rrrrrrrrr,
			head to column names = false,
			table head=\hline Ch & Runs & Peak & Conv & Lim & Iterations & Run Time & MaxErr & RMSD \\\hline,
			late after line=\\,
			late after last line=\\\hline]%
		{"ds1_convergence.csv"}%
		{}%
		{\csvcoli & \csvcolii & \csvcoliii & \csvcoliv & \csvcolv & \csvcolvi & \csvcolvii & \csvcolviii & \csvcolix}%
\label{TABdsoneconvergence}
\end{table}
}

\newcommand{\TABcover}{\
\begin{table}
\centering
\caption{The mean round trip cover time (MRT) for each dataset with and without MCMCMC.
}
	\csvreader[tabular=lrrr,
			head to column names = false,
			table head=\hline Data & 1-chain MRT & 4-chain MRT & Ratio \\\hline,
			late after line=\\,
			late after last line=\\\hline]%
		{"mean_cover_all.csv"}%
		{}%
		{\csvcoli & \csvcolii & \csvcoliii & \csvcoliv}%
\label{TABcover}
\end{table}
}

\newcommand{\TABtopogr}{\
\begin{table}
\centering
\caption{Estimated topology deviation (RMSD) and potential scale reduction factor (PSRF) using our topological Gelman-Rubin-like measure (TGR). Ch = number of chains.}
	\csvreader[tabular=lrrrrr,
			head to column names = false,
			table head=\hline Data & Ch & \multicolumn{2}{c}{TGR-RMSD} & \multicolumn{2}{c}{TGR-PSRF} \\
			 & & 2-runs & 8-runs & 2-runs & 8-runs \\\hline,
			late after line=\csvifoddrow{\\\hline}{\\},
			late after last line=\\\hline]%
		{"topo_gr.csv"}%
		{}%
		{\csvcoli & \csvcolii & \csvcoliii & \csvcolv & \csvcoliv & \csvcolvi}%
\label{TABtopogr}
\end{table}
}

\newcommand{\TABlegacytreebase}{\
\begin{table}
\centering
\caption{TreeBASE identifiers and legacy identifiers for the eukaryotic data sets used in this study.
}
	\csvreader[tabular=lll,
			head to column names = false,
			table head=\hline Data & ID & Legacy ID \\\hline,
			late after line=\\,
			late after last line=\\\hline]%
		{"legacy_treebase.csv"}%
		{}%
		{\csvcoli & \csvcolii & \csvcoliii}%
\label{TABlegacytreebase}
\end{table}
}

\newcommand{\TABtopoerror}{\
\begin{table}

\centering
\caption{The standard error of topology posterior probabilities for the top trees between the golden run eukaryote posteriors.
Columns labeled by posterior probabilities of the top trees.
Dashes from the left indicate that all posterior probabilities satisfied the given threshold.
Dashes from the right indicate that all posterior probabilities fit within the next smaller threshold.
}
	\csvreader[tabular=lcccccc,
			head to column names = false,
			table head=\hline Data & $<$ 1e-05 & $<$ 1e-04 &  $<$ 1e-03 &  $<$ 1e-02 &  $<$ 1e-01 &  $<$ 1e+00 \\\hline,
			late after line=\\,
			late after last line=\\\hline]%
		{"topo_error.csv"}%
		{}%
		{\csvcoli & \csvcoliv & \csvcolv & \csvcolvi & \csvcolvii & \csvcolviii & \csvcolix}%
\label{TABtopoerror}
\end{table}
}

\newcommand{\FIGspr}{\
\begin{figure}
\begin{center}
  \arxiv{\includegraphics[width=4in]{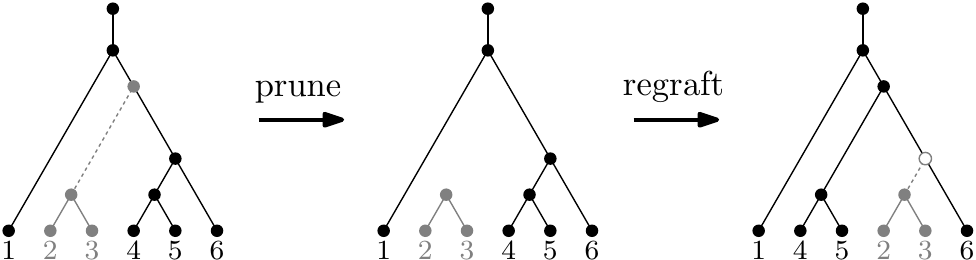}}
\end{center}
\caption{\
  An SPR move.
}
\label{FIGspr}
\end{figure}
}

\newcommand{\FIGusprcounterexample}{\
\begin{figure}
\begin{center}
  \arxiv{\includegraphics[width=4in]{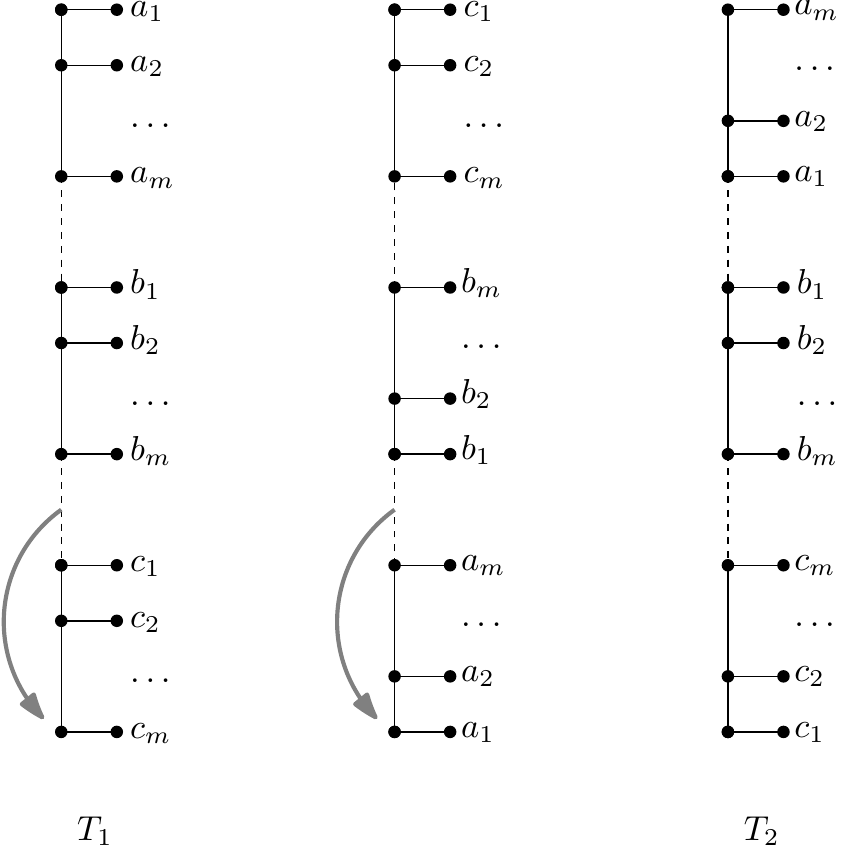}}
\end{center}
\caption{\
  Two trees such that the best rooting SPR distance overestimates the unrooted SPR distance.
	This occurs when every leaf is part of some moved subtree in every minimal unrooted set of SPR operations.
	In this example, the unrooted distance remains 2, while the best rooting SPR distance is $\lfloor\frac{m + 3}{2}\rfloor$ for $m \ge 4$.
	Dashed lines indicate the edges modified by these minimal unrooted SPR operations.
}
\label{FIGusprcounterexample}
\end{figure}
}

\newcommand{\FIGtopoerror}{\
\begin{figure}
\begin{center}
	\arxiv{\includegraphics[width=\textwidth]{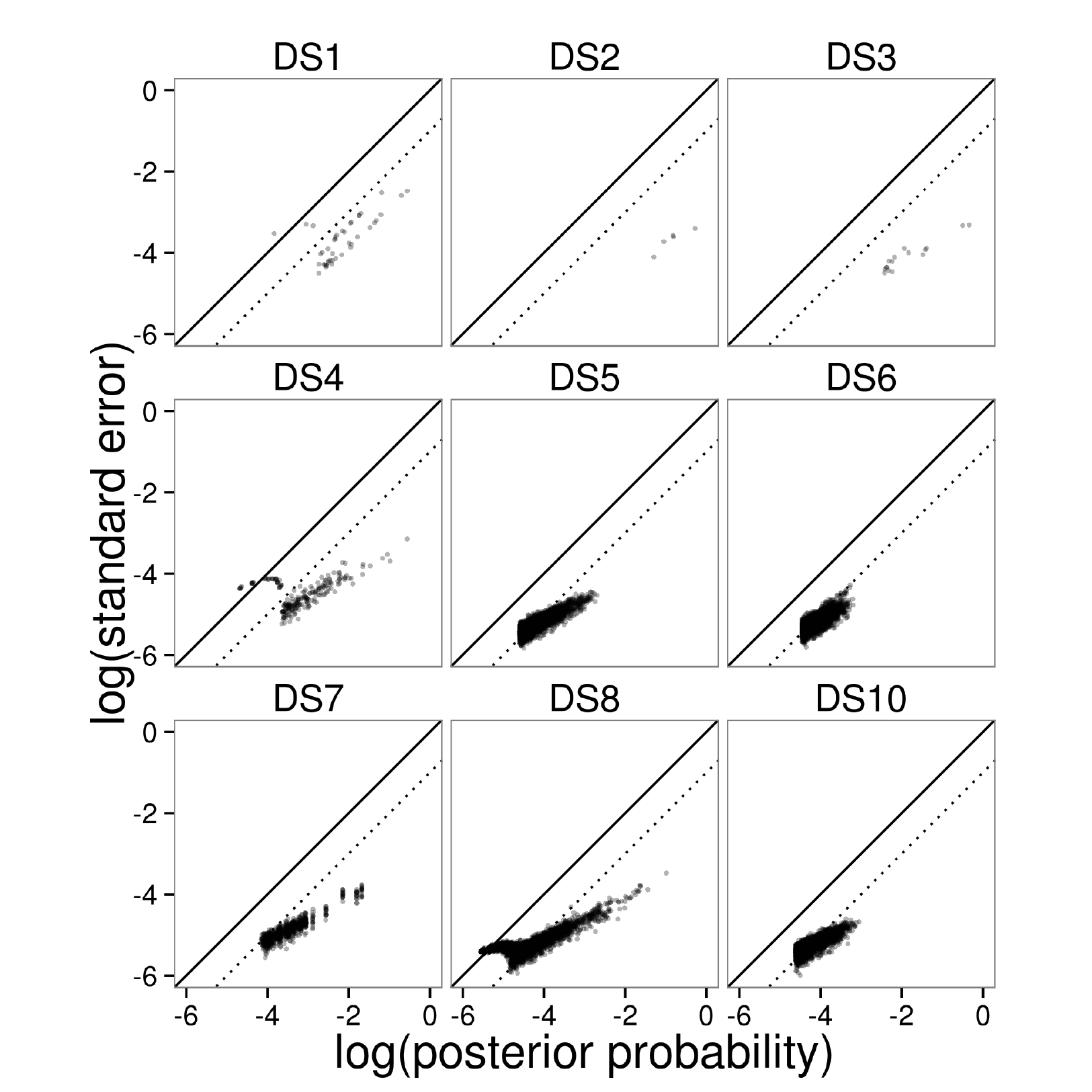}}
\end{center}
\caption{\
The standard error of topology posterior probabilities for the top trees between the golden run eukaryote posteriors.
The standard error is smaller than the posterior probability for estimates below the solid line and an order of magnitude smaller for estimates below the dotted line.
}
\label{FIGtopoerror}
\end{figure}
}

\newcommand{\FIGsprgraph}{\
\begin{figure}
\arxiv{
\centering
\includegraphics[width=0.8\textwidth]{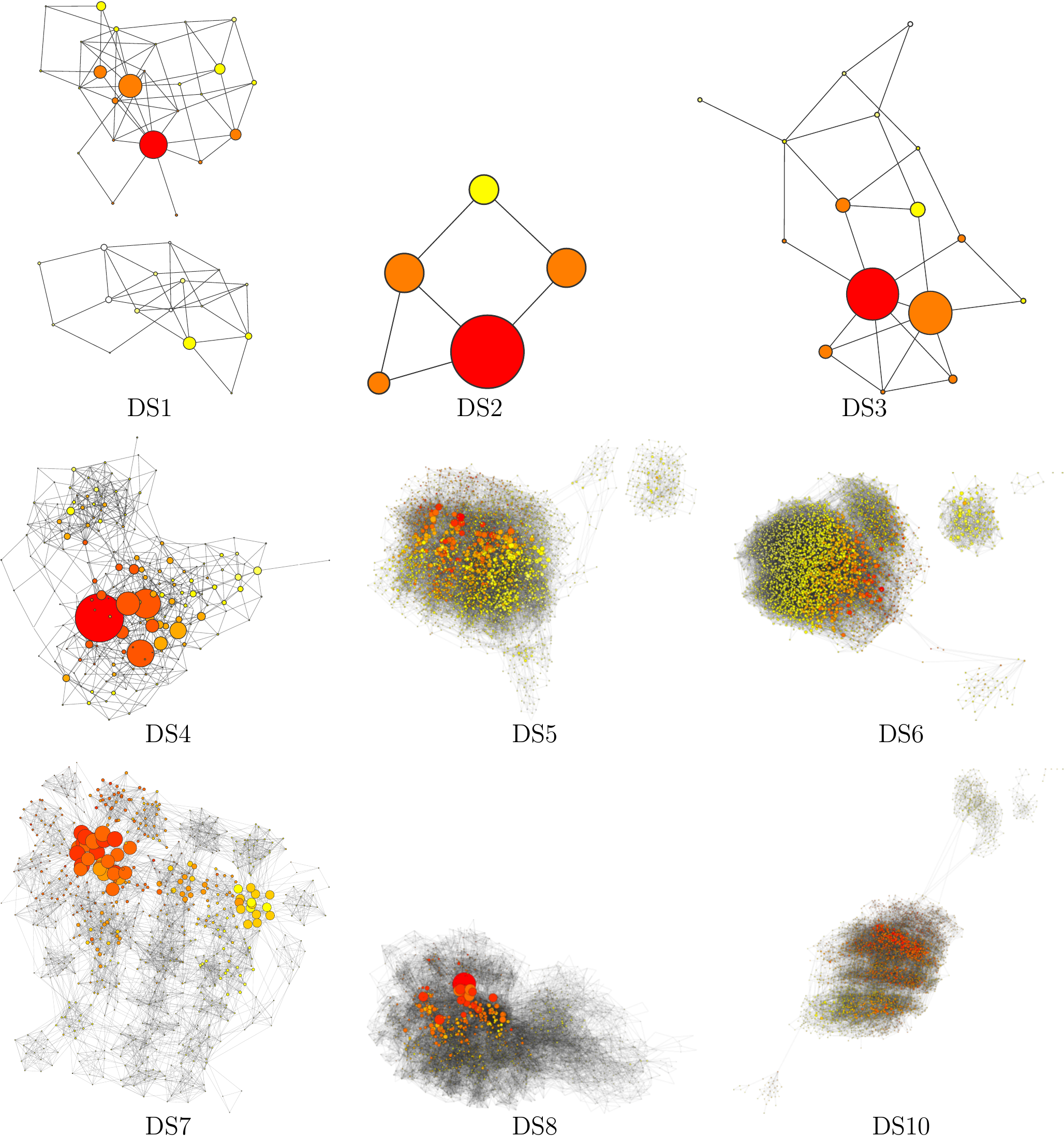}
}

\caption{Distance SPR graphs of the combined golden run eukaryote posteriors.
Each graph contains either the 95\% credible set or the 4096 topologies with highest estimated posterior probabilities (DS5, DS6 and DS10).
Node areas are scaled relative to posterior probability (PP; larger = higher probability) within each graph (but not with respect to the other graphs).
Node color indicates SPR distance from the topology with highest posterior probability in each dataset on a red-yellow-white scale (dark-light in the print version), with the highest probability tree colored red.
}
\label{FIGsprgraph}
\end{figure}
}

\newcommand{\FIGsprcluster}{\
\begin{figure}
\arxiv{
\centering
\includegraphics[width=\textwidth]{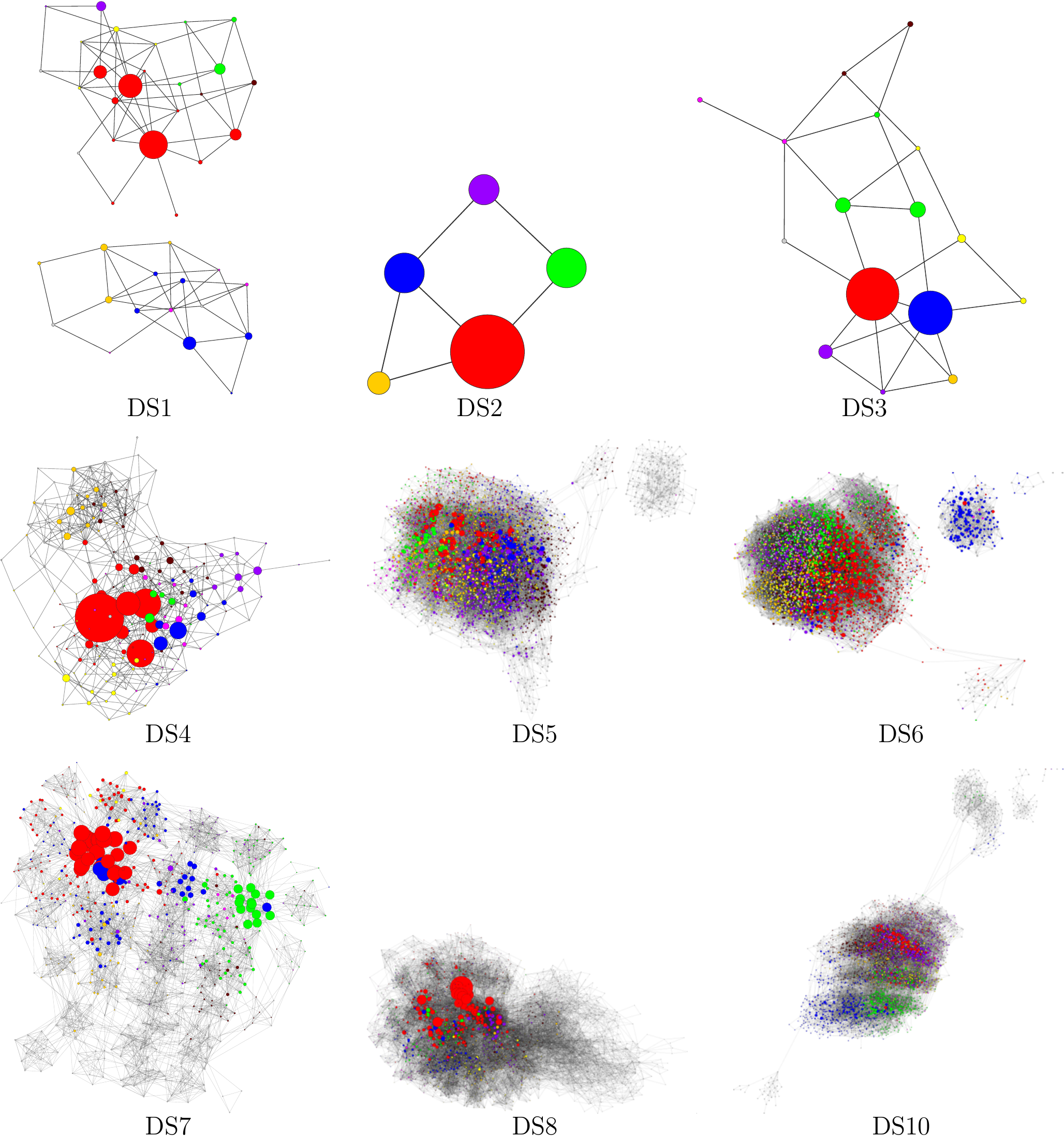}
}
\caption{Cluster SPR graphs of the combined golden run eukaryote posteriors.
Each graph contains either the 95\% credible set or the 4096 topologies with highest PP (DS5, DS6 and DS10).
Nodes are scaled relative to posterior probability within each graph (but not with respect to the other graphs).
Nodes are colored by SPR-based descending PP clusters (grayscale in the print version).
}
\label{FIGsprcluster}
\end{figure}
}

\newcommand{\FIGdsonepeaks}{\
\begin{figure}
\arxiv{
\centering
\includegraphics[width=\textwidth]{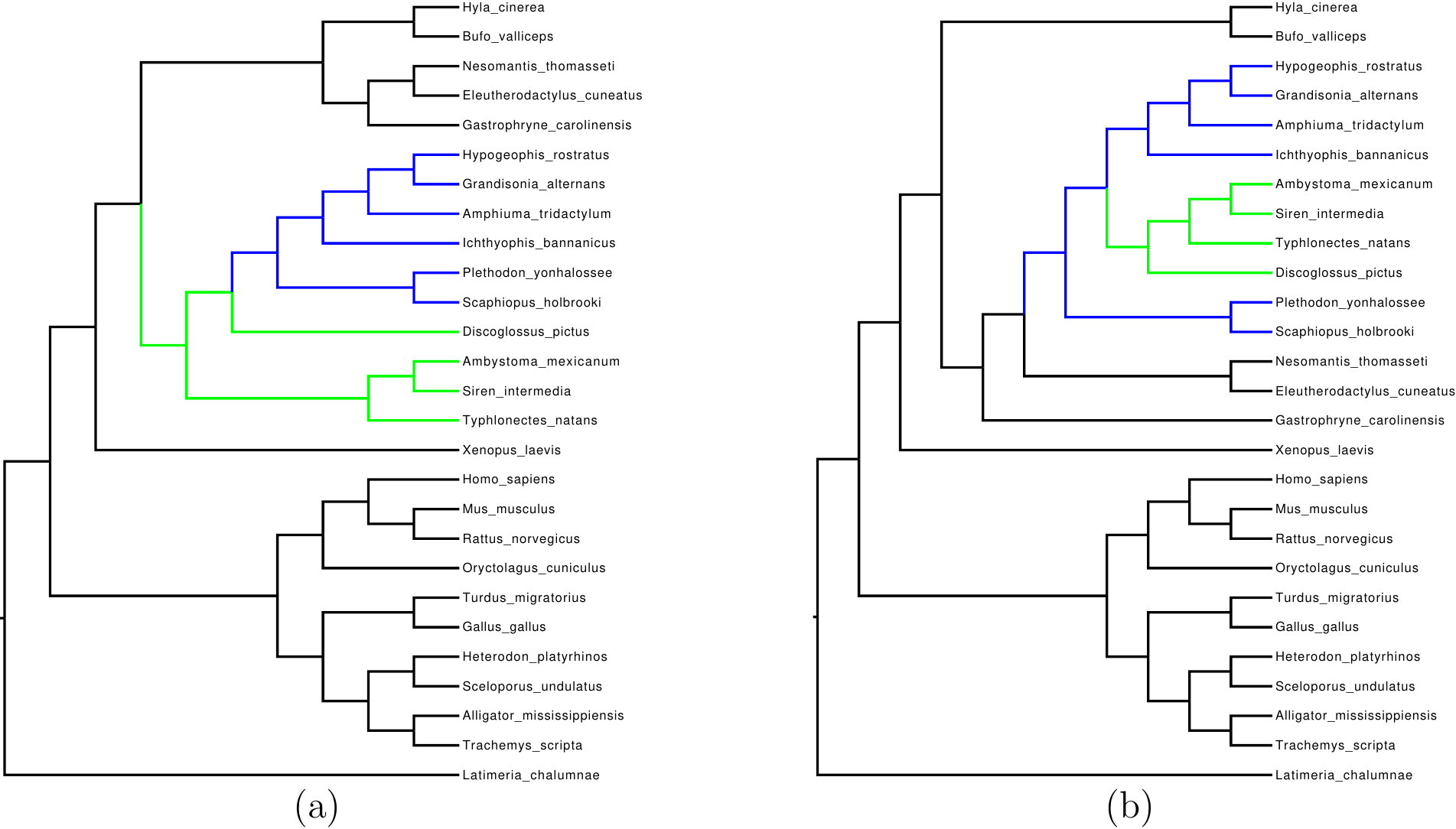}
}
\caption{Central trees of the two topological peaks in dataset DS1.
Only two SPR operations separate these trees, moving the blue (gray in the print version) and then green (light gray) clade to traverse from peak 1 to peak 2 and vice versa in the reverse direction.
However, the sole intermediate topology is so unlikely that it was never visited in any of our tests, inducing a severe topological bottleneck.
Longer paths through multiple trees outside of the 95\% confidence interval are taken instead, resulting in long transit times between the peaks.
}
\label{FIGdsonepeaks}
\end{figure}
}

\newcommand{\FIGdssevenunique}{\
\begin{figure}
\arxiv{
\centering
\includegraphics[width=0.75\textwidth]{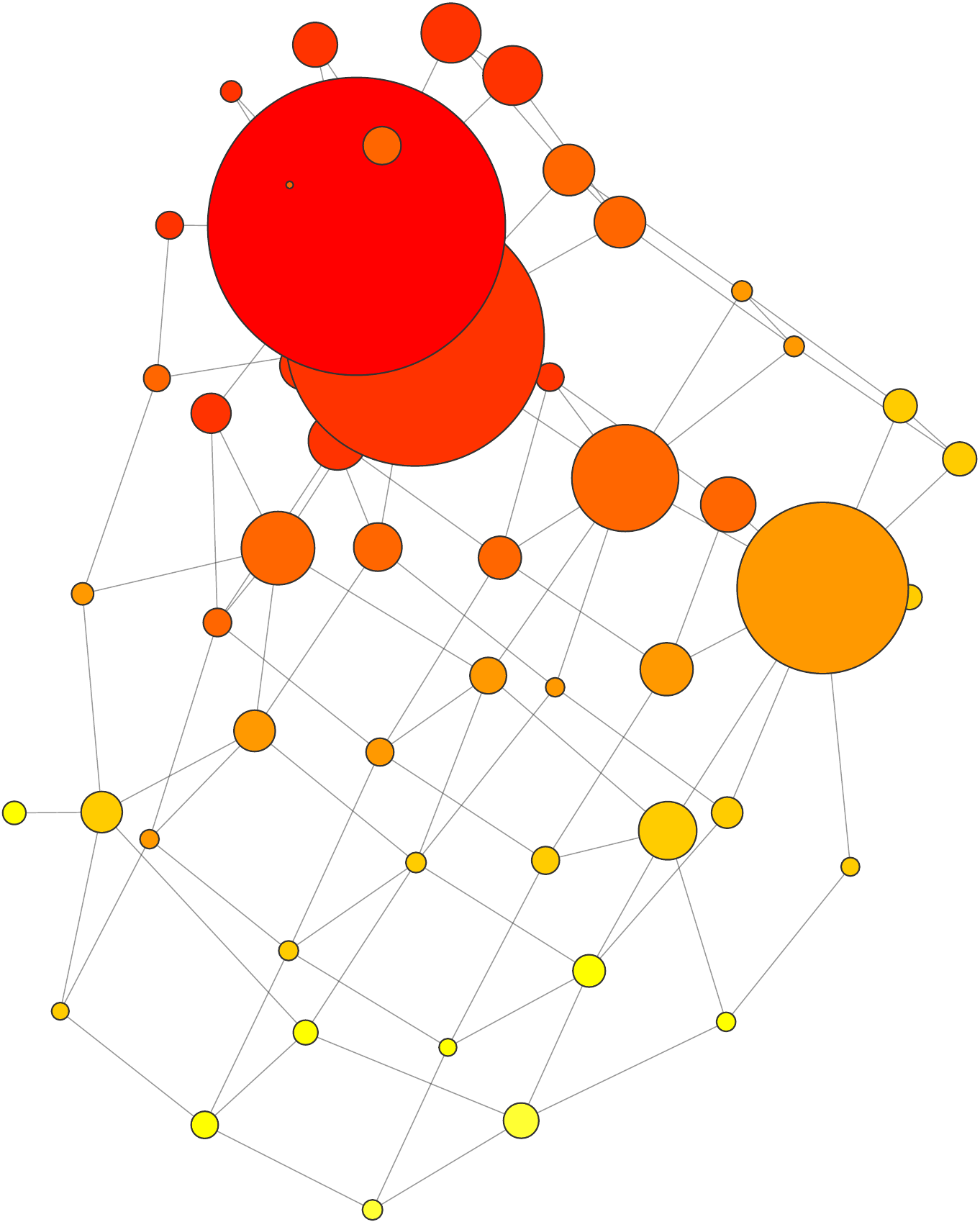}
}
\caption{Distance SPR graph of the 95\% credible set of the combined golden run eukaryote posteriors for DS7 with 3 of the 4 nearly identical \emph{Microcebus rufus} sequences removed.
Area indicates posterior probability and color indicates SPR distance from the topology with highest posterior probability on a red-yellow-white scale (dark-light in the print version), with the highest probability tree colored red.
}
\label{FIGdssevenunique}
\end{figure}
}

\newcommand{\FIGdssixpeaks}{\
\begin{figure}
\arxiv{
\centering
\includegraphics[width=\textwidth]{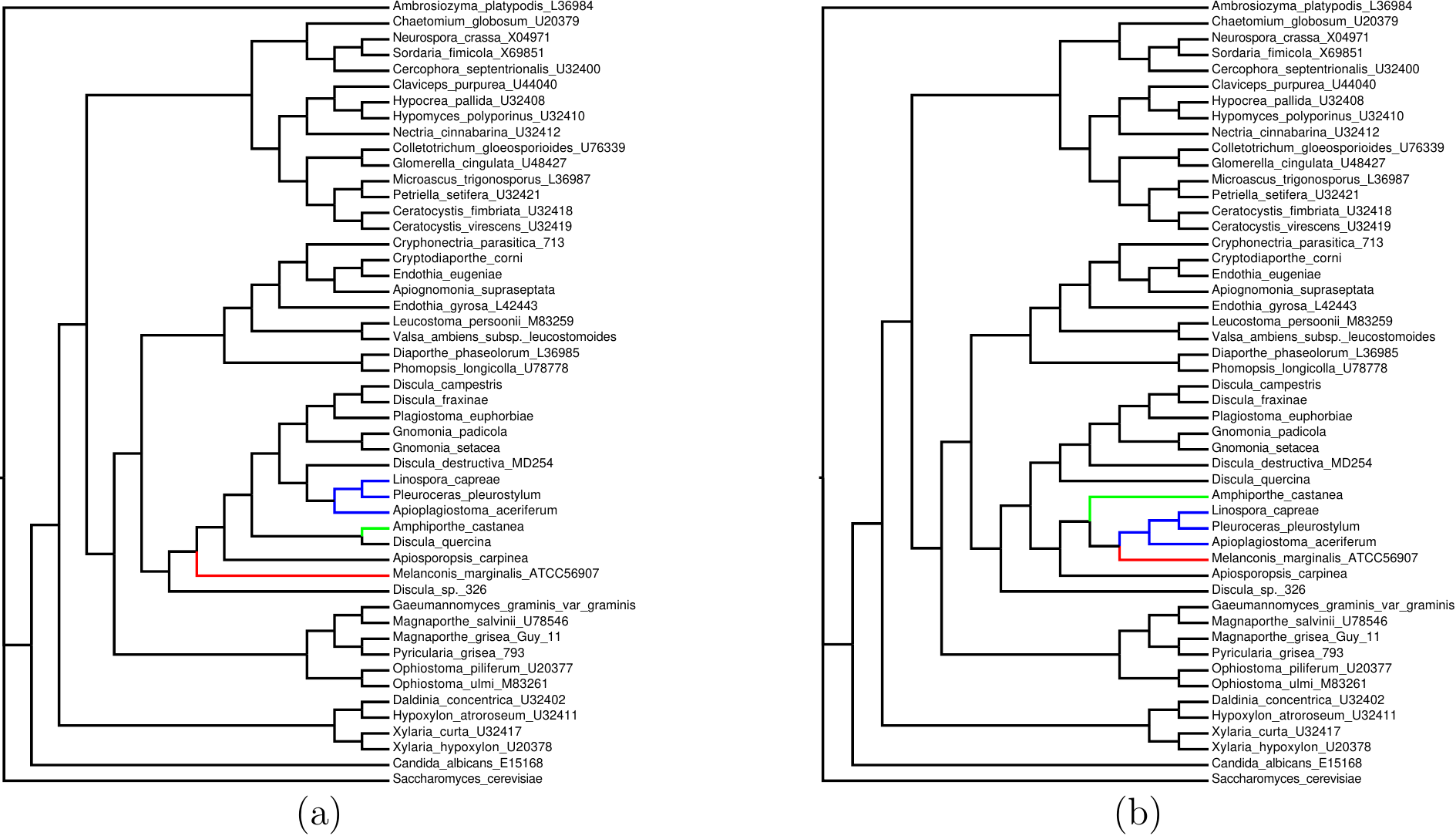}
}
\caption{Central trees of the two topological peaks in dataset DS6.
Only three SPR operations separate these trees, moving the colored subtrees.
Intermediate groupings are unsupported and intermediate trees unlikely.
}
\label{FIGdssixpeaks}
\end{figure}
}

\newcommand{\FIGdsoneconsensus}{\
\begin{figure}
\centering
\arxiv{\includegraphics[width=0.9\textwidth]{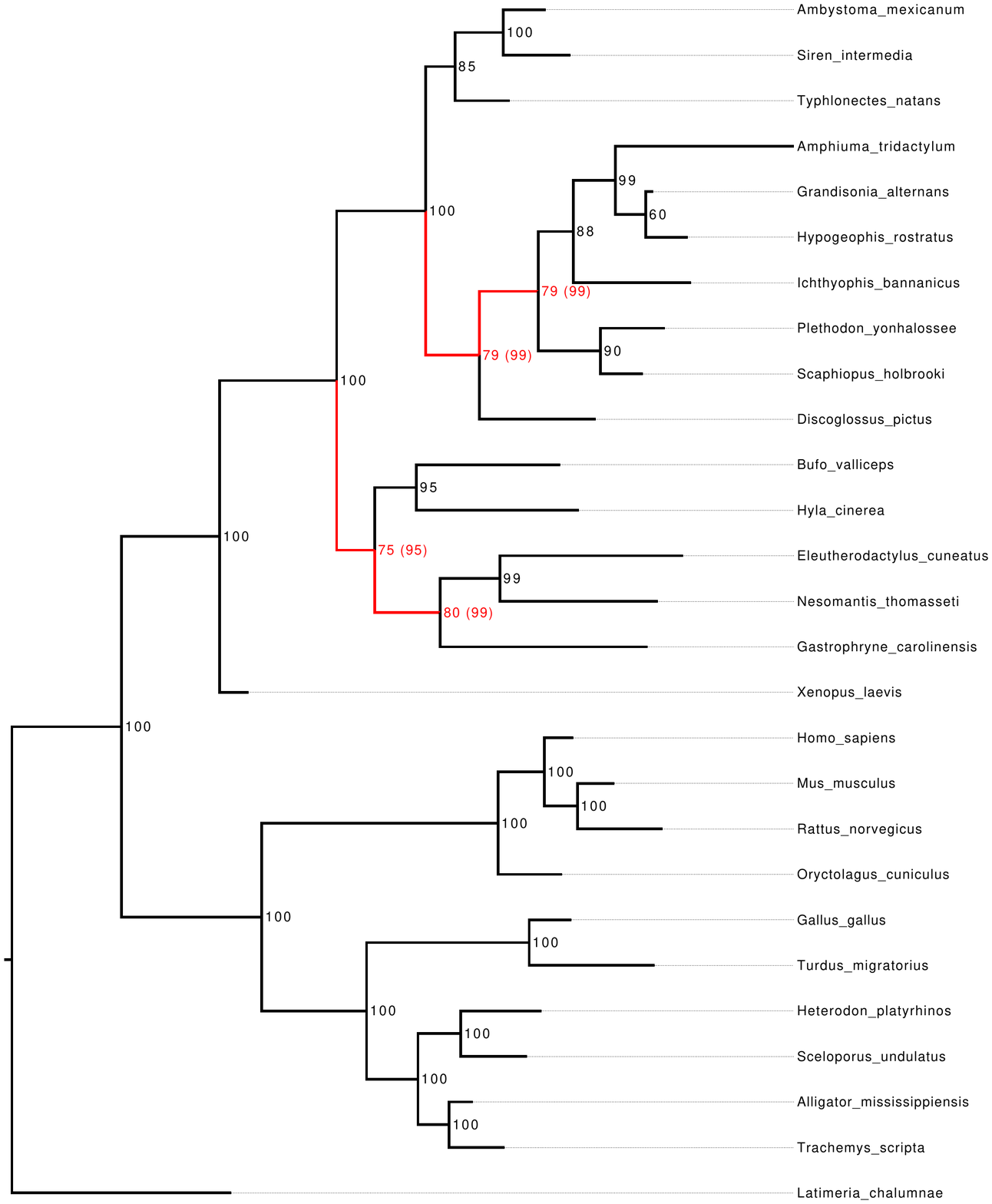}}
\caption{Extended Majority Rule Consensus tree of the DS1 golden runs.
Node labels indicate the percentage of trees within the 95\% credible set containing that split.
Four splits, indicated in red, receive erroneously high support when the second peak is not sampled (numbers in brackets)}
\label{FIGdsoneconsensus}
\end{figure}
}

\newcommand{\FIGweightedmcmc}{\
\begin{figure}
\arxiv{
\centering
\includegraphics[width=\textwidth]{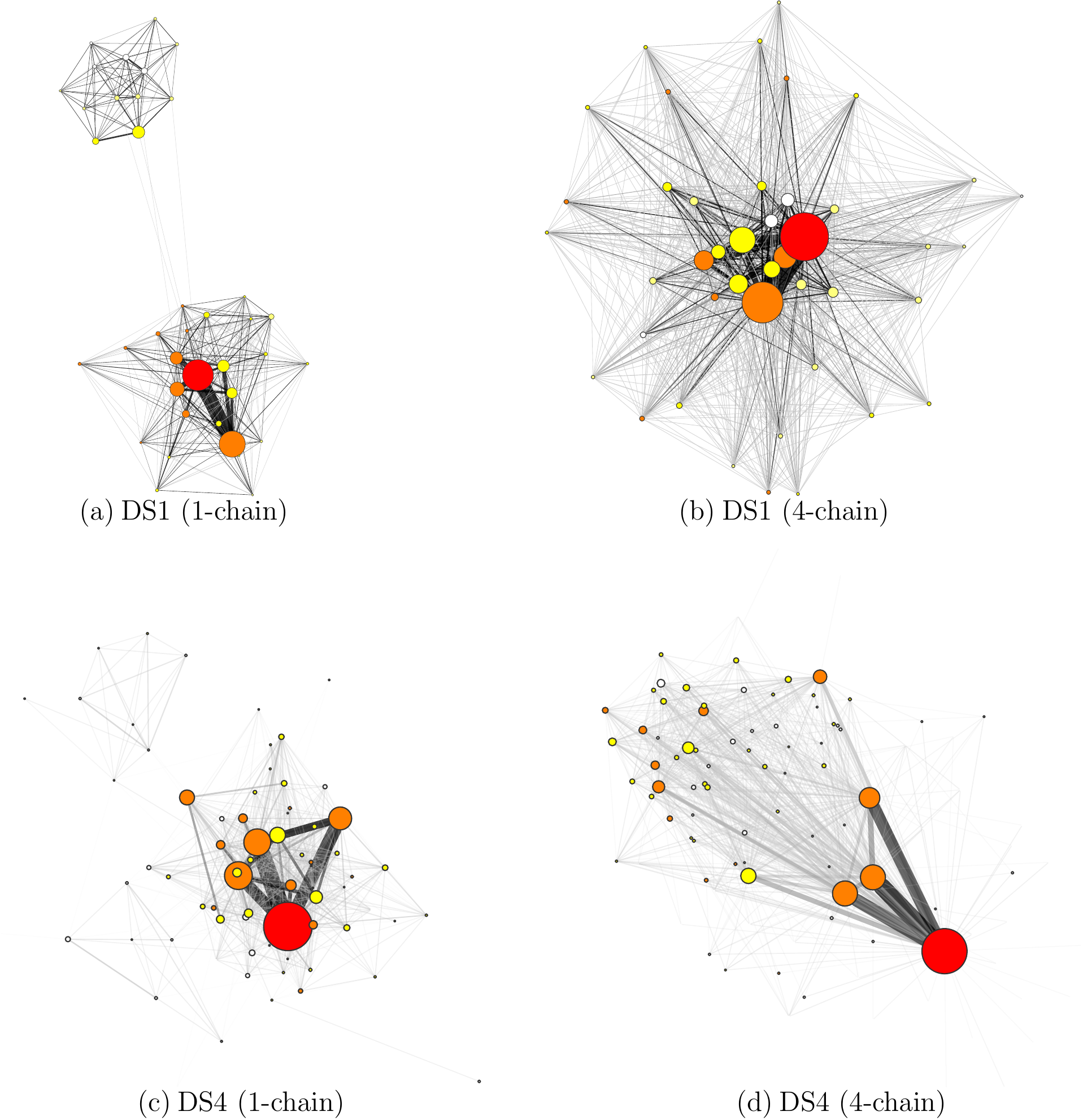}
}
\caption{Weighted MCMC graphs for DS1 and DS4.
Node diameters are scaled relative to posterior probability.
Nodes are colored on a red-yellow-white scale (dark-light in the print version) with increasing distance from the topology with highest posterior probability.
Edges connect trees in successive 100-iteration samples.
Edge thickness and color are proportional to the number of MCMC transitions.
}
\label{FIGweightedmcmc}
\end{figure}
}

\newcommand{\FIGmct}{\
\begin{figure}
\centering
\arxiv{\includegraphics[width=\textwidth]{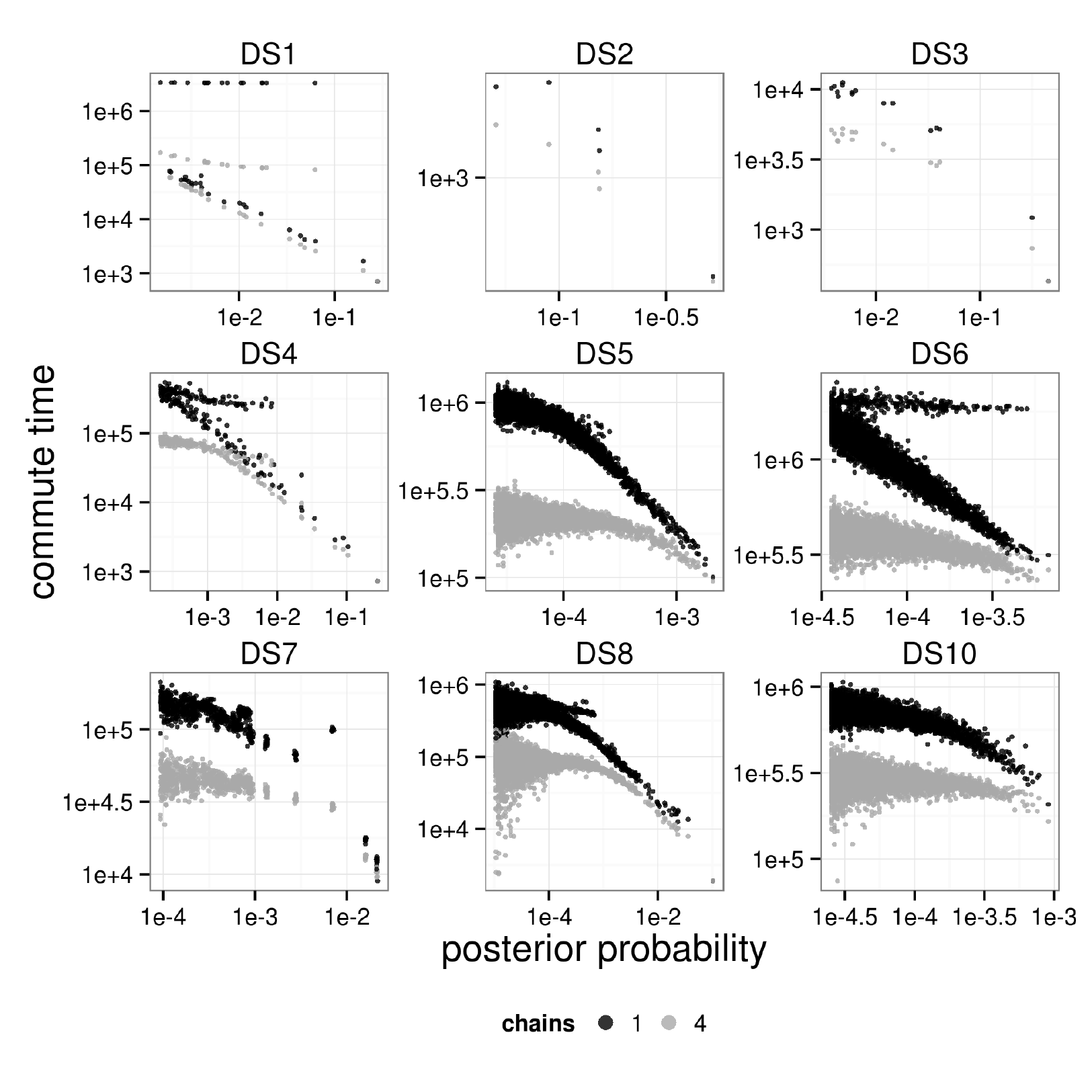}}

\caption{Comparison of posterior probability and mean commute time with (gray) and without (black) Metropolis-coupling.}
\label{FIGmct}
\end{figure}
}

\newcommand{\FIGmds}{\
\begin{figure}
\arxiv{
\centering
\includegraphics[width=0.7\textwidth]{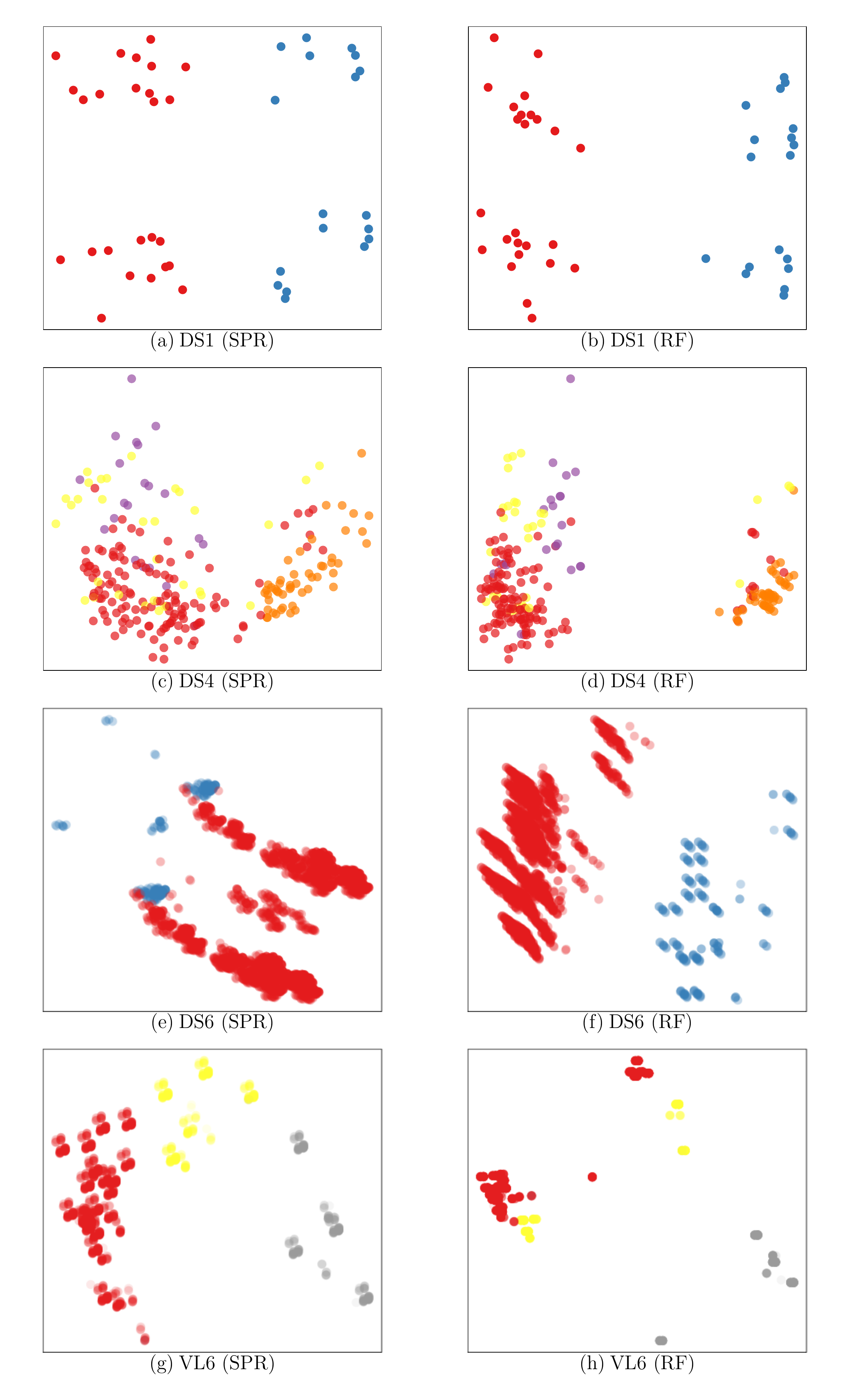}
}
\caption{Comparison of multidimensional scaling representations with SPR and RF distances.
Nodes are colored by identified peaks to match the primary cluster of the peak (grayscale in the print version).
}
\label{FIGmds}
\end{figure}
}

\newcommand{\FIGmdsflat}{\
\begin{figure}
\arxiv{
\centering
\hspace*{\stretch{1}}%
\includegraphics[width=\textwidth]{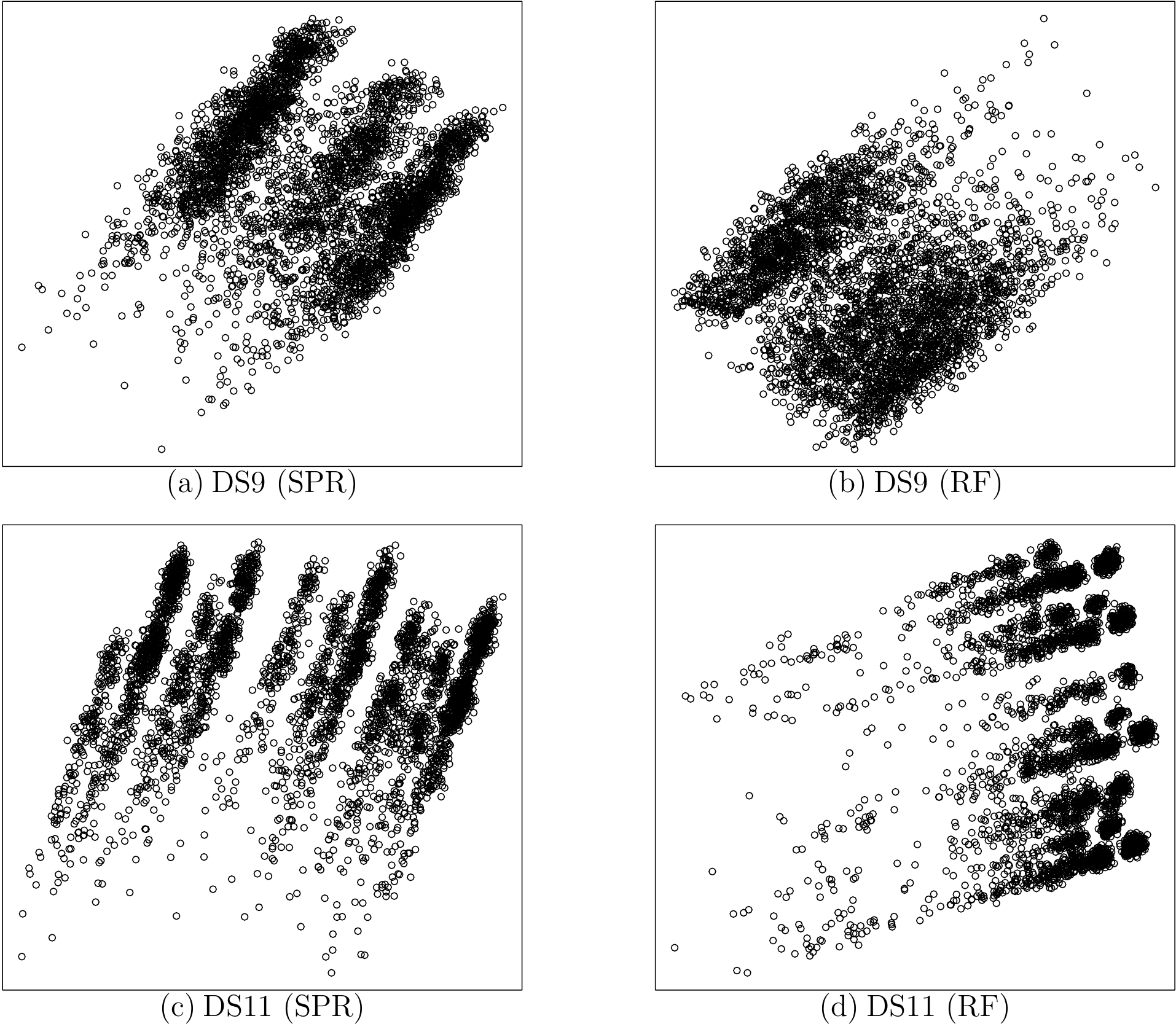}
}
\caption{Comparison of multidimensional scaling representations with SPR and RF distances for flat posteriors DS9 and DS11.
}
\label{FIGmdsflat}
\end{figure}
}

\newcommand{\FIGbeikograph}{\
\begin{figure}
\arxiv{
\centering
\includegraphics[width=0.8\textwidth]{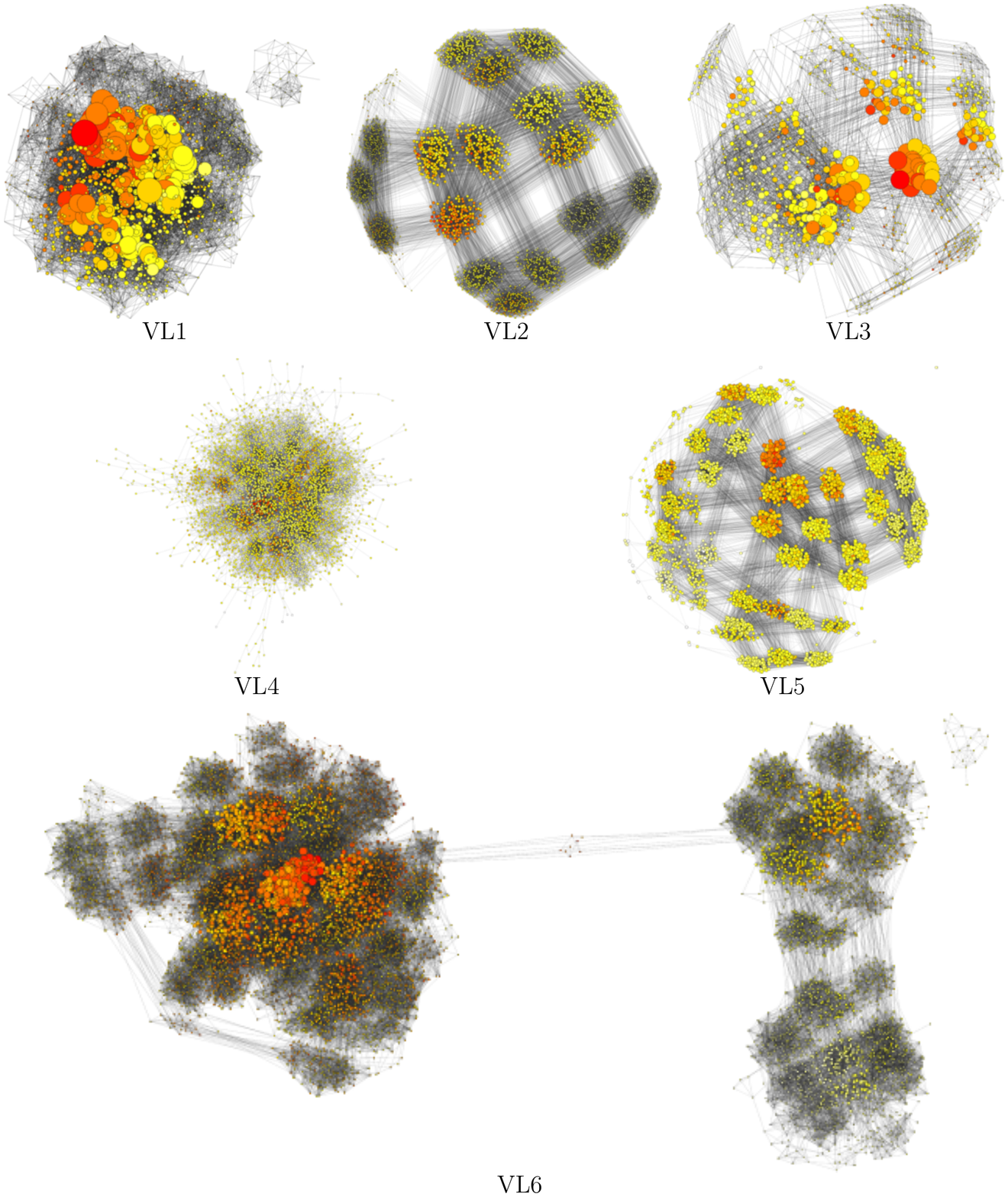}
}
\caption{Distance SPR graphs of the combined bacterial and archaeal golden runs showing at most the 4096 topologies with highest posterior probability (8192 for VL6).
Node areas are scaled relative to posterior probability (PP; larger = higher probability) within each graph (but not with respect to the other graphs).
Node color indicates SPR distance from the topology with highest posterior probability in each dataset on a red-yellow-white scale (dark-light in the print version), with the highest probability tree colored red.
}
\label{FIGbeikograph}
\end{figure}
}

\newcommand{\FIGerror}{\
\begin{figure}
\centering
\arxiv{\includegraphics[width=\textwidth]{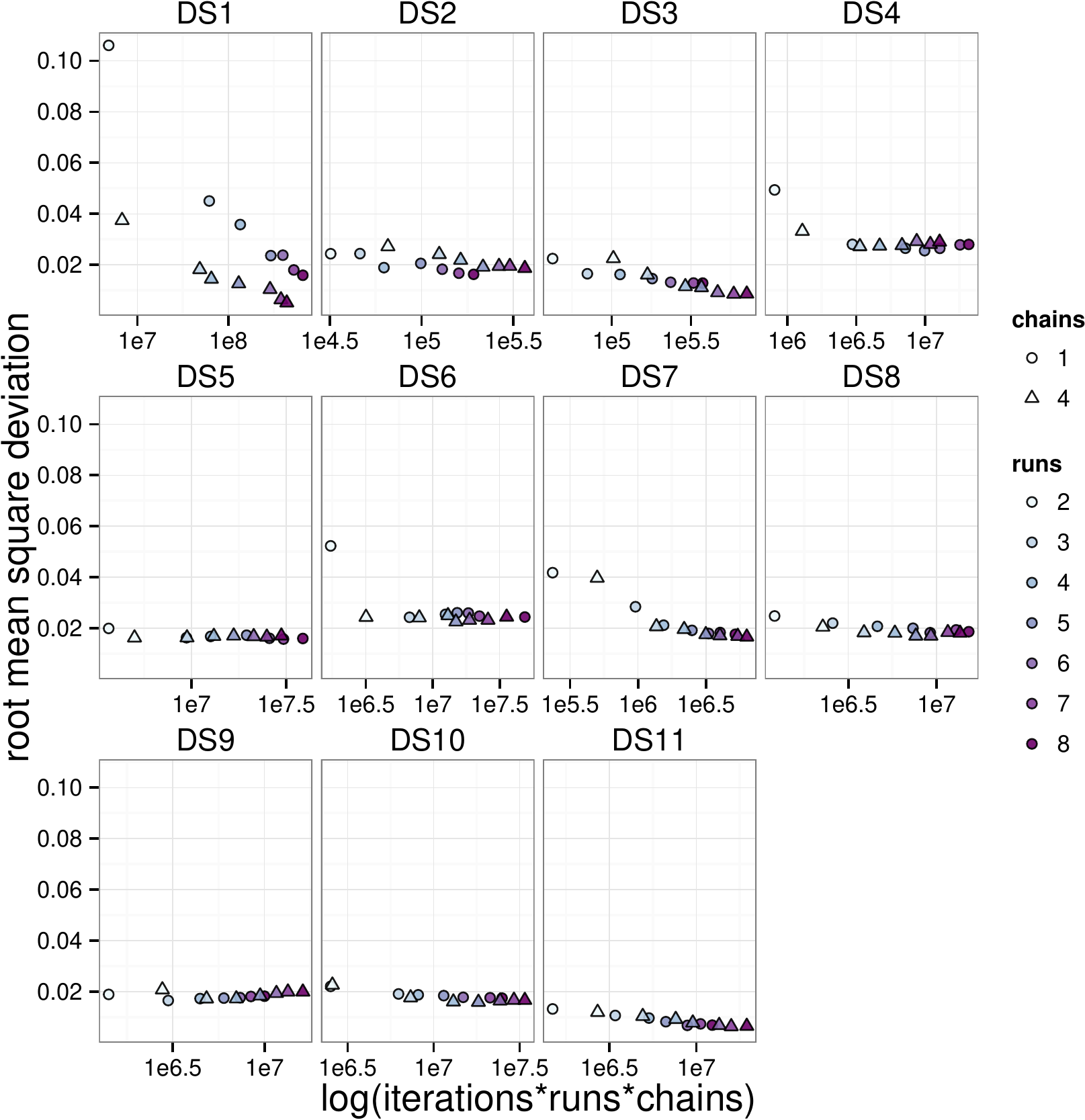}}
\caption{Comparison of mean running time and split frequency error with and without Metropolis-coupling using varying numbers of runs.}
\label{FIGerror}
\end{figure}
}

\newcommand{\FIGccp}{\
\begin{figure}
\arxiv{
\centering
\includegraphics[width=\textwidth]{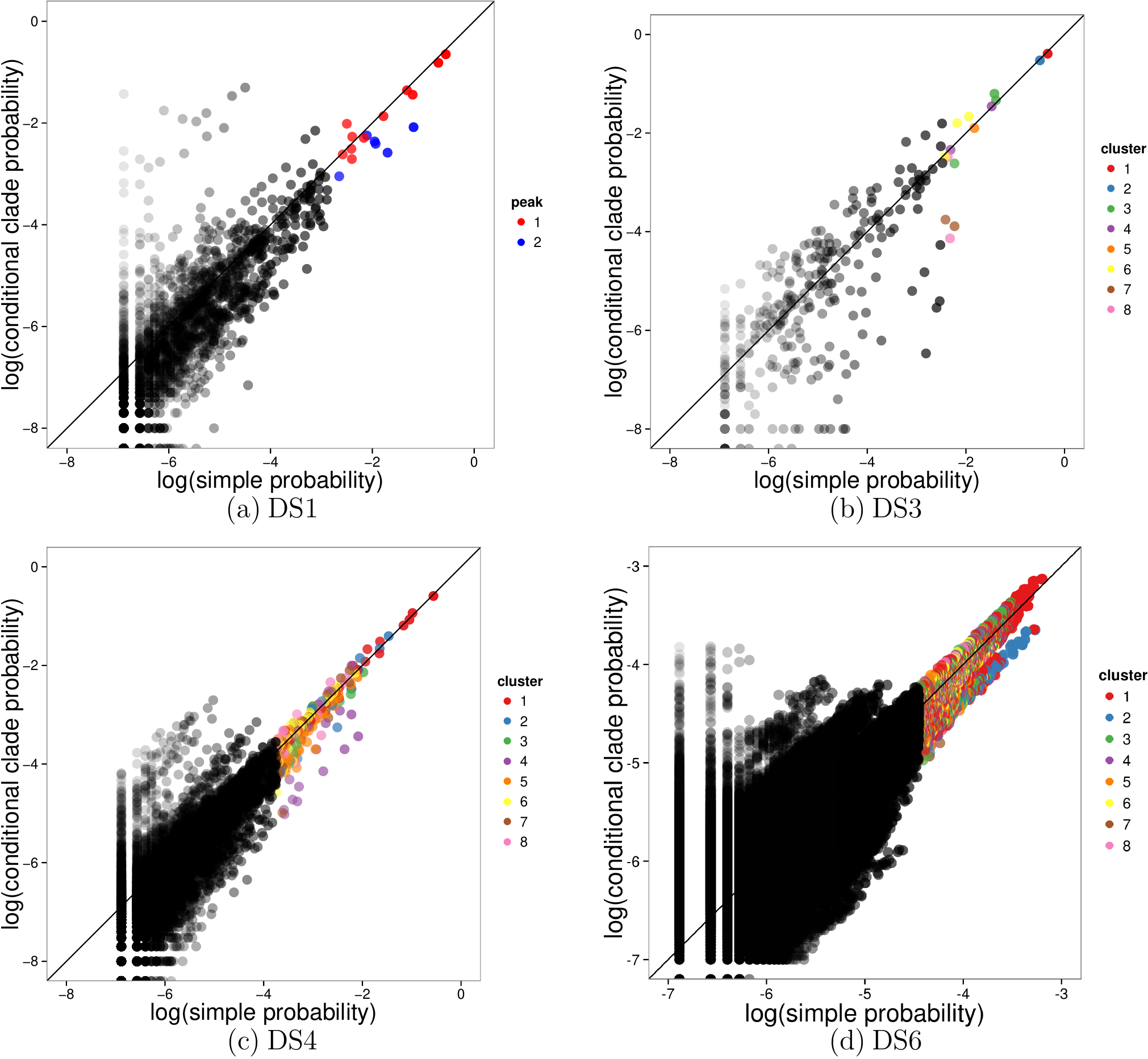}
}
\caption{A comparison of posterior probability and CCD estimates for the aggregated golden runs on datasets DS1, DS3, DS4, and DS6.
Probability is shown on a log-log scale in base 10.
The top trees for each dataset are colored by peak in DS1 and cluster for the other datasets.
Transparency of points increases as posterior probability decreases.
}
\label{FIGccp}
\end{figure}
}

\newcommand{\FIGtopogr}{\
\begin{figure}
\arxiv{
\centering
\includegraphics[width=\textwidth]{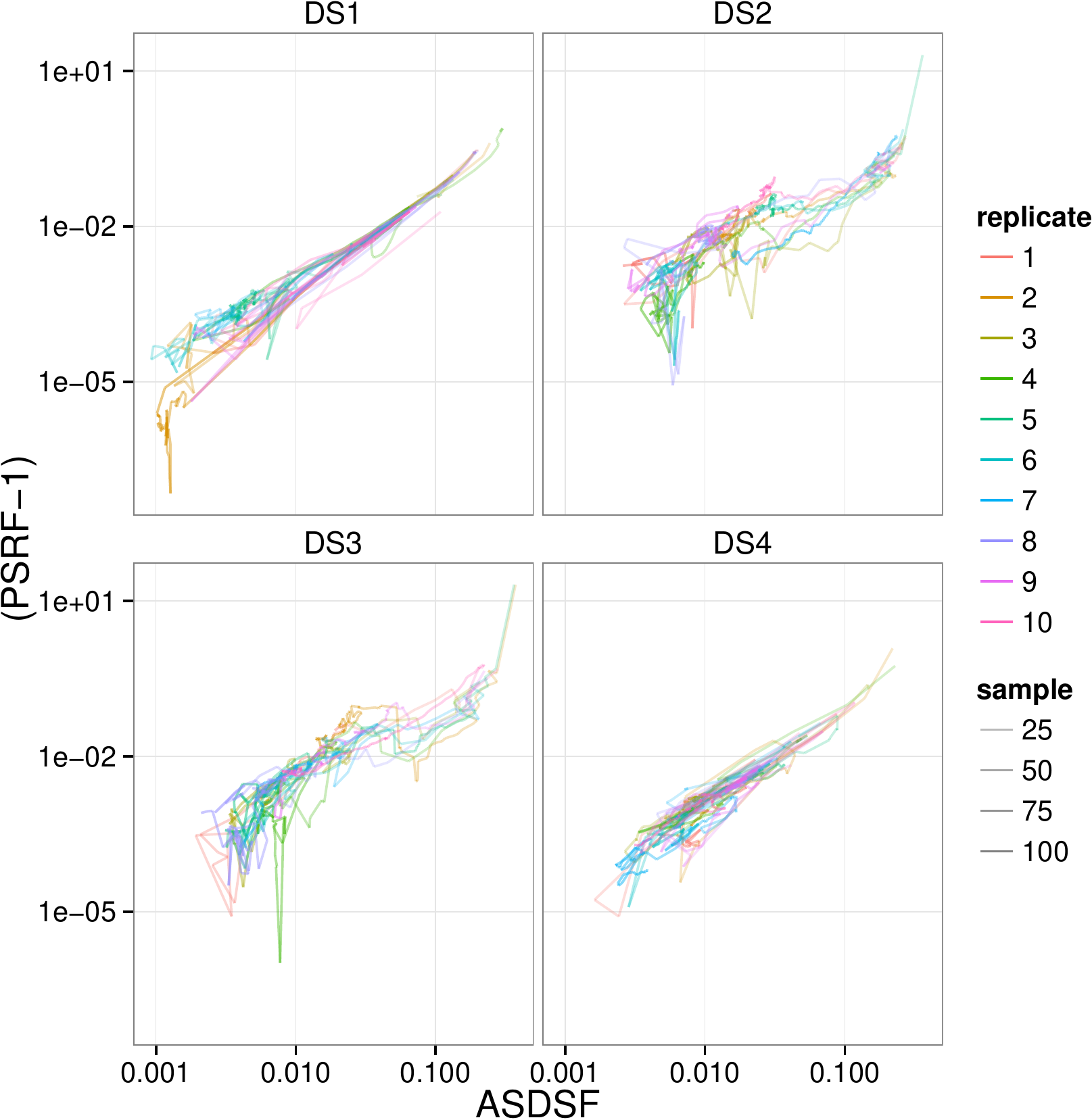}
}
\caption{A comparison of the average standard deviation of split frequencies and SPR topological Gelman-Rubin-like convergence diagnostics for datasets DS1, DS2, DS3, and DS4 using 2 independent runs.
Values are shown on a log-log scale in base 10.
100 evenly-spaced samples were taken from the first 2 runs of each 8-run replicate that had achieved ASDSF of less than 0.01.
Transparency decreases with time.
}
\label{FIGtopogr}
\end{figure}
}

\renewcommand{\section}[1]{%
\bigskip
\begin{center}
\begin{Large}
\normalfont\scshape #1
\medskip
\end{Large}
\end{center}}
\renewcommand{\subsection}[1]{%
\bigskip
\begin{center}
\begin{large}
\normalfont\itshape #1
\end{large}
\end{center}}
\renewcommand{\subsubsection}[1]{%
\vspace{2ex}
\noindent
\textit{#1.}---}
\renewcommand{\tableofcontents}{}
\bibpunct{(}{)}{;}{a}{}{,}  
\raggedright
\renewcommand\bibsection{\section{\refname}}

\begin{document}

\notarxiv{\
\begin{flushright}
Version dated: \today
\end{flushright}
\bigskip
\noindent RH: QUANTIFYING MCMC EXPLORATION OF TREE SPACE
\bigskip
\medskip
\begin{center}

\noindent{\Large \bf Quantifying MCMC exploration of phylogenetic tree space.}
\bigskip

\noindent {\normalsize \sc
Chris Whidden$^1$ and Frederick A. Matsen IV$^1$}\\
\noindent {\small \it
$^1$
Program in Computational Biology, Fred Hutchinson Cancer Research Center, Seattle, WA, 98109, USA}\\
\end{center}
\medskip
\noindent{\bf Corresponding author:} Chris Whidden, Program in Computational Biology, Fred Hutchinson Cancer Research Center, Seattle, WA, 91802, USA; E-mail: cwhidden@fhcrc.org.\\
\vspace{0.5in}
}
\arxiv{
\title{Quantifying MCMC exploration of phylogenetic tree space}
\author{Christopher Whidden and Frederick A. Matsen IV}
\date{\today}
\maketitle
\begin{abstract}
}%

\notarxiv{
\subsubsection{Abstract}
}
In order to gain an understanding of the effectiveness of phylogenetic Markov chain Monte Carlo (MCMC), it is important to understand how quickly the empirical distribution of the MCMC converges to the posterior distribution.
In this paper we investigate this problem on phylogenetic tree topologies with a metric that is especially well suited to the task: the subtree prune-and-regraft (SPR) metric.
This metric directly corresponds to the minimum number of MCMC rearrangements required to move between trees in common phylogenetic MCMC implementations.
We develop a novel graph-based approach to analyze tree posteriors and find that the SPR metric is much more informative than simpler metrics that are unrelated to MCMC moves.
In doing so we show conclusively that topological peaks do occur in Bayesian phylogenetic posteriors from real data sets as sampled with standard MCMC approaches, investigate the efficiency of Metropolis-coupled MCMC (MCMCMC) in traversing the valleys between peaks, and show that conditional clade distribution (CCD) can have systematic problems when there are multiple peaks.

\arxiv{
\end{abstract}
\pagebreak
}

\notarxiv{
\vspace{1in}
\noindent (Keywords: Markov chain Monte Carlo; phylogenetic methods; subtree prune-and-regraft; topological peaks; tree space)\\
\newpage
}

The Bayesian paradigm has been extensively adopted to infer phylogenetic trees and associated parameter values in a consistent probabilistic framework.
[We are interested in convergence properties on the discrete structure of unrooted tree topologies, so for the purposes of this paper we will use the word \emph{tree} without further qualification to signify an unrooted leaf-labeled tree topology without branch lengths.]
Current Bayesian phylogenetic methods rely on being able to move efficiently through tree hypothesis space with a random walk via Markov chain Monte Carlo (MCMC)~\citep{metropolis1953equation,hastings1970monte}.
These include the widely used BEAST~\citep{drummond2007beast,drummond2012bayesian,bouckaert2014beast} and MrBayes~\citep{Ronquist2012-hi} software packages as well as more recent methods such as BAli-Phy~\citep{suchard2006bali}, RevBayes (\url{http://github.com/revbayes/revbayes}) and ExaBayes (\url{http://sco.h-its.org/exelixis/web/software/exabayes/index.html}).
The empirical distribution of suitably spaced MCMC samples converges to its true posterior distribution given an infinitely long run of the MCMC~\citep[reviewed in][]{tierney1994markov}.
However, in order to obtain accurate computations of trees and associated confidence levels in practice, it is essential that these Markov chains explore phylogenetic ``tree space'' efficiently.

Many important questions remain unanswered concerning the practical performance of MCMC for phylogenetics, such as the presence and frequency of multiple peaks (i.e.\ modes) in phylogenetic tree posteriors, and the ability of chains to move between these posterior peaks.
Also, to what extent are ``peaky'' (i.e.\ multimodal) posteriors a consequence of the discrete structure of phylogenetic trees, or simply a consequence of simultaneously estimating a large number of real parameters?
Do strategies such as Metropolis-coupled MCMC (MCMCMC or (MC)$^3$)~\citep{geyer1992practical,huelsenbeck2001mrbayes}, which are helpful for multimodal distributions in the real case, effectively solve the problem?
To what extent do convergence diagnostics based on tree topologies, such as average standard deviation of split frequencies between independent Markov chains, imply that the empirical distribution of the underlying discrete tree topologies is close to the actual posterior?
How many independent chains are required for such convergence diagnostics to adequately assess the level of convergence?

There are continuing \citep{lakner2008efficiency,vstefankovic2011fast,hohna2012guided} and sometimes vitriolic \citep{mossel2005phylogenetic,ronquist2006comment,mossel2006limitations} debates concerning how well MCMC methods explore tree space.
\citet{lakner2008efficiency} and \citet{hohna2008clock} showed that the random choices of operations used in current methods lead to a low rate of accepted transitions and increase the amount of computation required before MCMC runs achieve a given split frequency distance to golden runs.
To address this problem, \citet{hohna2012guided} introduced improved Metropolized Gibbs samplers---biased operators that use additional computation to select transitions with a higher acceptance rate---and showed that these operators reduced the time to achieve such a given split frequency distance to golden runs using BEAST on 11 empirical data sets.
Parsimony-biased tree proposals have been included in MrBayes 3.2 \citep{Ronquist2012-hi}.
\citet{mossel2005phylogenetic} showed mathematically that MCMC methods can give misleading results when the alignments used to construct the trees derive from a site-wise mixture of data generated on two very different trees (note that this usage of ``mixture'' refers to a means of combining probability distributions, whereas the separate concept of ``mixing'' as described below refers to a characteristic of Markov chains).
On such a site-wise mixture, the Markov chain appears to converge rapidly according to diagnostics but in actuality requires an exponential amount of time to converge due to the large ``valleys'' of unlikely trees between the two site-wise mixture peaks.
Such site-wise mixtures are but one contrived example of a peaky distribution.
However, even if we never see the sort of data set they postulate we may still encounter peaky distributions.
In such a situation, the posterior samples from a single peak may appear as though the chain has completely explored the relevant part of tree space, leading to a mistakenly high confidence value for an incomplete sample of trees.
Although there has been extensive discussion in the literature about to what extent Metropolis-coupling helps traverse peaks, there have been few conclusions, probably because there hasn't been a clear exploration of peaks and peakiness in phylogenetic posteriors.

Some studies have focused on estimating mixing properties of phylogenetic MCMC using theory~\citep{aldous2000mixing,mossel2005phylogenetic}; this is known to be a very hard problem and can only be done in ``toy'' examples.
[As is standard in the field, we will use the word \emph{mixing} to refer to the convergence of the empirical distribution of MCMC samples to their posterior distribution.]
Even when we can diagnose the failure of a Markov chain to converge to the posterior distribution, it does not lead to an understanding of why the failure occurred.
A more practical approach to understanding movement in discrete tree space is to equip this space with a metric and consider distances traveled by the chain.

Recently, \citet{hohna2012guided} and \citet{larget2013estimation} proposed using Conditional Clade Probability (CCP) and Conditional Clade Distribution (CCD) methods, respectively, to approximate tree posterior probabilities.
In both methods, the probability of a tree is estimated based on a product of conditional clade probabilities.
\citet{larget2013estimation} uses the approximation that compatible splits, separated by another split, are approximately conditionally independent given the separating split.
The approximating equation of CCD is then a product of joint conditional sister clade probabilities, given the parent clade.
Conditional probability methods have the potential to estimate the posterior probabilities of many trees using only a small sample of the tree posterior.
They have already been productively applied to approximate tree posteriors in phylogenomic analyses \citep{szollosi2013efficient}.
However, the validity of the assumption of conditional independence of sister clades, given the parent clade, is not clear in practice.
It is thus crucial to determine the accuracy of the CCD approximation on real data sets.

\arxiv{\FIGspr}
These considerations motivate improved methods to understand the performance of phylogenetic methods and the corresponding ``topography'' of trees.
\citet{hillis2005analysis} used the Robinson-Foulds (RF) distance~\citep{robinson81} between phylogenies with multidimensional scaling (MDS) to visualize tree space.
However, the RF distance does not correspond to SPR operators, and in fact may be arbitrarily large even for trees separated by a single SPR operation.
\citet{matsen2006geometric} suggested using the nearest-neighbour interchange (NNI) distance with MDS visualization.
\citet{hohna2012guided} used this idea to visualize ``islands'' among 15 trees from the 27 taxon tree space.
Still, the NNI distance does not correspond closely with many rearrangements used in phylogenetic inference and is difficult to compute, limiting the utility of this method.

Subtree prune-and-regraft (SPR)~\citep{hein96} moves are the most common rearrangements used by phylogenetic programs~\citep{hohna2012guided}.
These involve cutting a subtree off and attaching it somewhere else (Fig.~\ref{FIGspr}).
The minimum number of such operations required to transform one tree into another is called the SPR distance.
Moreover, SPR operators are closely related to other common rearrangements.
NNI operators are a subset of SPR operators.
Two other common operators, the subtree swap (SS) and tree-bisection-and-reconnection (TBR) are each equivalent to two SPR operations \citep{hohna2012guided}.

Thus the SPR distance is especially appropriate to investigate phylogenetic MCMC behaviour in this setting because of the correspondence between SPR operators and most MCMC moves.
However, SPR distance is challenging to use due to the computational complexity of its computation~\citep{allen01,bordewich05,hickey2008sdc}.
Recently, efficient fixed-parameter algorithms for computing the SPR distance have been developed and implemented in the freely available and open source RSPR software package~\citep{whidden2009unifying,whidden2010fast,whidden2013hybridization,whidden2014supertrees}.
These efficient algorithms require fractions of a second to compute SPR distances between trees with hundreds of taxa and enable, for the first time, tree comparison using the SPR distance on a relevant scale.

In this paper, we use SPR tree space to visualize and analyze Bayesian phylogenetic posterior distributions.
Our graph-based method directly shows the difficulty in moving between areas of tree space and can identify topological peaks that are not visible in multidimensional scaling projections.
We show that our SPR graphs explain the error rate and time to a given average standard deviation of split frequencies (ASDSF)~\citep{Ronquist2012-hi} of Bayesian phylogenetic methods on various data sets when these statistics do not correlate with the number of taxa alone.
Moreover, we show that multiple topological peaks are common in nontrivial posteriors, even with relatively few taxa, and that the graphs can be used to identify bottlenecks in posterior distributions: regions of tree space between peaks that are difficult for MCMC methods to cross.
We propose a topological variant of the Gelman-Rubin convergence diagnostic and show that a small ASDSF often implies a small such topological convergence diagnostic.
We explore the effect of Metropolis-coupling and show that it greatly improves mixing, particularly between topological peaks, and reduces the number of MCMC iterations required for multiple runs to achieve a given ASDSF threshold.
Metropolis-coupling improves overall performance in peaky distributions but may increase computation time in non-peaky distributions, in which case we observe the number of iterations to be reduced by a smaller factor than the number of Metropolis-coupled chains.
For both MCMCMC and single-chain approaches, we find that the current standard of two runs to calculate ASDSF is insufficient to obtain a proper error estimate.
Finally, we show that independence of sister clades, conditioned on parent clades, does not hold in some peaky distributions.
This causes the CCD distribution to systematically underestimate the probability of trees within alternative peaks and systematically overestimate the probability of trees between peaks.

\section{Methods}

\subsection{Computing the SPR distance}

We modified RSPR, the open source C++ software package for computing subtree prune-and-regraft distances~\citep{whidden2010fast,whidden2013hybridization,whidden2014supertrees}.
Previous versions of RSPR computed the SPR distance between two input trees or the aggregate SPR distance from a single tree to a set of trees.
Our new version 1.3 of RSPR (\url{https://github.com/cwhidden/rspr/}) adds support for computing pairwise SPR and RF distance matrices.
These distance matrices can be used as input to multidimensional scaling methods or to compute tree space graphs.

RSPR computes a maximum agreement forest (MAF)~\citep{hein96,allen01} of two rooted trees with a fixed-parameter algorithm.
An agreement forest is a forest of subtrees that can be obtained by cutting edges from both trees.
An MAF is obtained by cutting the fewest possible number of edges.
This smallest number of cut edges is equivalent to the SPR distance between the trees if they are rooted \citep{bordewich05}.
The time required for this fixed-parameter algorithm increases exponentially with the distance computed but only linearly with the size of the trees.
In particular, the algorithm can quickly determine whether two rooted trees are separated by a single SPR operation.
In practice, RSPR can compute SPR distances between trees with hundreds of taxa and more than 50 transfers in fractions of a second~\citep{whidden2014supertrees}.

Unrooted trees are commonly inferred by phylogenetic methods including MrBayes.
However, an MAF of two unrooted trees is equivalent to their tree-bisection-and-reconnection distance \citep{allen01} and no MAF formulation is known for the SPR distance of unrooted trees.
For unrooted trees, we thus consider each possible rooting of the trees and choose the rootings which give the minimal SPR distance.
This ``best rooting'' SPR distance should closely agree with the unrooted SPR distance except in pathological cases where the minimum set of unrooted SPR operations is incompatible with any rooting (e.g. Supplemental Figure~\ref{FIGusprcounterexample}).
In particular, both are guaranteed to agree when the trees are separated by a single SPR operation; much of our work here uses the graph induced by these single SPR moves.

\subsection{SPR tree space graphs}
\label{sec:graph}
We used SPR-based graphs, restricted to sets of high probability trees, to model the SPR tree space of Bayesian phylogenetic posterior distributions.
We selected these sets of high probability trees as follows.
First, we ordered the trees from a posterior sample by descending posterior probability (ties broken by sample order).
In cases with a large number of ties (e.g. where every tree is sampled once or twice), breaking ties with sample order may cause bias, so we broke ties randomly in such cases.
The 95\% credible set is the smallest set of trees at the head of this list with cumulative posterior probability more than 95\%.
We call the $m$ trees with highest posterior probability the ``top $m$ trees,'' that is, the first $m$ trees in this list.
We used $m=4096$ in our tests unless otherwise noted, and generally used the 95\% credible set when it contained fewer than 4096 trees, and the top 4096 trees when it was not.
We call these sets of at most 4096 trees the ``top trees''.

We define the SPR graph for a set of trees $T$ to be the undirected graph $G_T = (V, E)$ such that each tree is represented by a node in $V$ and two trees are connected by an edge in $E$ if and only if they are separated by an SPR distance of 1.
In particular, we constructed a distance matrix $D$ such that an entry $D_{ij}=1$ if, and only if, the SPR distance between $i$ and $j$ is 1.
We constructed such graphs using RSPR version 1.3, then converted these matrices to an edge list format suitable for input to graph visualization software.

\subsection{Clustering high-probability regions of tree space}
\label{sec:cluster}

We used a simple iterative clustering procedure to aid in the detection of topological peaks.
These peaks are intuitively defined as a set of topologies with relatively high probability surrounded by topologies with low probability.
Any useful clustering procedure must therefore make use of posterior probabilities in addition to topology, moreover, comparing every pair of trees is computationally expensive even with the simple goal of computing RF distances.
We thus employed the following approximate iterative clustering algorithm.
First select the most probable topology as the center of our first cluster.
Then compare the current cluster center to each unclustered tree, and add each tree within a specified SPR distance radius to the current cluster.
This procedure proceeds iteratively, grouping the most probable unclustered topology and the remaining set of unclustered trees until each tree has been clustered or a given number of clusters assigned.
For a given cluster center, we used a clustering radius equal to the mean SPR distance from the current cluster center to each unclustered tree, minus the standard deviation of these distances (i.e. $\mu - \sigma$).
This radius is recalculated for each new cluster.
We stopped this process after 8 clusters had been identified.

\subsection{Graph visualization with Cytoscape}
\label{sec:cytoscape}

SPR graphs were visualized with the open source Cytoscape platform~\citep{shannon2003cytoscape}.
In addition to the edge list and clusters described above, we computed SPR distances between the tree with highest posterior probability and the top $m$ trees.
We visualized tree space in three ways: (1) distance SPR graphs, (2) cluster SPR graphs, and (3) weighted MCMC graphs.
To visualize SPR graphs we used a force-directed graph layout, which essentially means that graph nodes are pushed away from each other, but edges act as ``springs'' that attempt to maintain a uniform length.
We scaled node sizes (area) in proportion to tree posterior probability.
The largest node represents the tree with highest posterior probability.
We hypothesized that peaks would be visible in such graphs as sets of relatively large (high probability) nodes separated by relatively small (low probability) nodes or in disconnected graph components.
In distance SPR graphs, graph nodes are colored on a red-yellow-white scale (dark-light in the print version) with increasing SPR distance from the most probable topology.
We further hypothesized that difficult to sample peaks would be visible in distance SPR graphs as large yellow or white nodes.
In clustered SPR graphs, graph nodes are colored by cluster.
We expected that any significant topological peaks would be grouped in different clusters and therefore receive different colors.
Finally, we used another type of graph to visualize Markov chain movement between trees to validate our assumption that SPR tree space corresponds to MCMC movement in practice.
These graphs represent movement between MCMC samples (including Metropolis-coupling chain swaps where applicable).
We weighted these edges with the number of such transitions and visualized these edge weights using edge thickness and color.
Note, however, that posteriors are typically subsampled every given number of iterations, and we followed this practice.
Given such subsamples, some of the dependence between sample tree and order may be eliminated and care must be taken when interpreting such graphs.

\subsection{Quantifying tree space mixing}
\label{sec:commute}

To quantify mixing behaviour in tree space we computed statistics based on mean access times (MAT)---the mean number of iterations required to transition between topologies in an MCMC search \citep{lovasz1993random}.
As with our graph clustering, computing statistics for each pair of trees can be computationally expensive and difficult to visualize.
Rather than directly considering access time statistics for each pair of trees, we instead computed the mean commute time (MCT) \citep{lovasz1993random} from the most probable topology to each other high probability tree and back: the sum of pairwise MATs.
We also considered a new measure, the mean round trip cover time.
This is the mean number of iterations required to cover (visit) each high probability tree, starting from and returning to the highest probability tree.
This measure is essentially a round-trip analog of the mean cover time \citep{lovasz1993random}.
The MAT values (and hence MCT and round trip cover time values) can be computed with a single pass through the tree posterior using a method for updating weighted means \citep[see e.g.][]{west1979updating}. Formal definitions of these statistics and a description of our dynamic programming method for computing them can be found in the supplementary material.

\subsection{A discrete topological Gelman-Rubin-like convergence diagnostic}

In order to avoid having to project trees down to vectors of split frequencies in order to diagnose convergence, we developed a discrete topological variant of the Gelman-Rubin convergence diagnostic \citep{gelman1992inference}.
The Gelman-Rubin convergence diagnostic for a \textit{real-valued} parameter $x$ requires multiple independent Markov chains and compares the variance within chains and between chains, as we review now.
Note that by ``chains'' here we refer to multiple independent chains, which are equivalent to the MrBayes terminology ``runs'' rather than Metropolis-coupled chains.
Suppose we have $m$ chains, each with $n$ sampled values.
The value of chain $i$ at iteration $j$ is denoted $x_{ij}$.
The variance between chains, $B$, is estimated by the variance between the $m$ sequence means, $\bar{x}_{i.}$, each based on $n$ values of $x$.
That, is,
$$B/n = \frac{1}{m-1} \sum_{i=1}^m (\bar{x}_{i \cdot} - \bar{x}_{\cdot \cdot})^2,$$
where
$\bar{x}_{\cdot \cdot} =  \frac{1}{m} \sum_{i=1}^m \bar{x}_{i \cdot}$\,.
The variance within chains, $W$, is the average of the $m$ within-sequence variances, $s_i^2$, each based on $n-1$ degrees of freedom.
That is,
$$W = \frac{1}{m}\sum_{i=1}^m s_i^2,$$
where
$ s_i^2 = \frac{1}{n-1} \sum_{j=1}^n (x_{ij} - \bar{x}_i)^2.$
The estimated variance is then a weighted average of $W$ and $B$,
$$\hat{V} = \left(1 - \frac{1}{n}\right) W + \frac{1}{n} B.$$
The potential scale reduction factor (PSRF) is defined as
$\hat{R} = \sqrt{\hat{V} / W}$.
This measures the potential for reducing the difference between $B$ and $W$.
$B$ initially overestimates the variance, given multiple chains with overdispersed starting points.
$W$ initially underestimates the variance, as it is based on an incomplete sample from a limited region of the parameter space.
These values converge as the independent chains converge.
As such, the PSRF approaches 1 as the chains converge.

Our topological Gelman-Rubin-like convergence diagnostic estimates the differences within and between Markov chains in terms of topological changes.
There is no concept of sample mean for topologies, so we compute an analogous statistic with the mean square deviation instead of variance.
In particular, we estimate the SPR distance deviation within and between chains.
Again, $x_{ij}$ denotes the tree from chain $i$ at iteration $j$.
Let $d(x_{i_1j_1},x_{i_2j_2})$ denote the distance between two such trees.

$W$ is the mean square deviation within a chain:
$$ s_i^2 = \frac{1}{n (n-1) } \sum_{j_1=1}^n \sum_{j_2=1}^n d(x_{ij_1}, x_{ij_2})^2.$$
Similarly, we estimated the between-chain deviation by comparing each chain to the aggregate set of chains :
$$B = \frac{1}{(m-1)mn^2} \sum_{i_1=1}^m \sum_{i_2=1}^m \sum_{j_1=1}^n \sum_{j_2=1}^n d(x_{i_1j_1}, x_{i_2j_2})^2.$$
With this formulation, $\sqrt{V}$ estimates the topology root mean square deviation (RMSD).
$\hat R$ is computed as before.

As written, these formulas require a great deal of computation.
To efficiently compute topological PSRF values, observe that there are many repeated comparisons between identical trees.
We thus grouped identical topology comparisons, computed one SPR distance for each and weighted the squared distances accordingly in our calculations.
We also limited our comparisons to the top trees, as in our SPR graph construction.
We normalized our computations by the number of included distances rather than the total number of samples $n$.
Using this method, $B$ is no more complex to compute than $W$.

As with the original Gelman-Rubin convergence diagnostic, the topological PSRF value approaches 1 as the independent chains converge.
B initially overestimates the RMSD between topologies, given multiple chains with overdispersed starting points.
W initially underestimates the RMSD between topologies, as it is based on an incomplete sample from a limited region of tree space.
These values converge as the independent chains converge.
We use the name topological Gelman-Rubin\emph{-like} to emphasize that it is inspired by the original but is not the same.

\label{sec:topogr}
\subsection{Multidimensional scaling}

Multidimensional scaling (MDS) is a method for projecting complex data to a small number of dimensions suitable for visualization \citep{kruskal1964multidimensional,kruskal1964nonmetric}.
Non-metric MDS is typically applied to create a new two or three dimensional space from a given pairwise distance matrix in a way that preserves the pairwise distances as much as possible.
Specifically, it minimizes a \textit{stress function} quantifying the difference between the original distances and Euclidean distances in the projected space.
Multidimensional scaling has been used previously to visualize RF distances between trees in a posterior distribution \citep{hillis2005analysis} and, on a limited scale, NNI distances \citep{hohna2012guided}.
We applied MDS to SPR and RF distance matrices using the R \texttt{isomds} function from the MASS package \citep{venables2002mass}.

\subsection{Conditional clade probability}

Recently, the conditional clade probability (CCP) and conditional clade distribution (CCD) concepts have been proposed by \citet{hohna2012guided} and \citet{larget2013estimation}, respectively.
These methods use conditional products of split posterior probabilities on splits to estimate the corresponding phylogenetic posterior probabilities.
To test the conditional independence assumption in practice, we applied the CCD software of \citet{larget2013estimation} to compute conditional clade probabilities and compare the results to posterior probabilities on large posterior samples.

\subsection{Number of runs and chains}

Two MrBayes run parameters are of particular importance to obtain ASDSF estimates that reflect the level of convergence to the posterior distribution: the number of independent runs used for testing ASDSF convergence and the number of Metropolis-coupling chains.
The number of independent runs determines the behavior of the average standard deviation of split frequencies (ASDSF) convergence diagnostic, which compares split frequencies between independent runs.
As is typical, our ASDSF calculations only consider splits with a frequency exceeding 10\% in at least one of the runs.
We follow previous researchers by using a 0.01 cutoff for ASDSF as a stopping rule.
In the MrBayes version 3.2 manual, \citet{ronquist2011draft} suggest that ``an average standard deviation below 0.01 is very good indication of convergence, while values between 0.01 and 0.05 may be adequate depending on the purpose of your analysis.''
Increasing the number of runs increases the stringency of ASDSF convergence at a given limit at the expense of increased computation.
Metropolis-coupling \citep{geyer1992practical,huelsenbeck2001mrbayes} is a commonly applied method to improve MCMC mixing in peaky distributions.
In addition to the primary ``cold'' Markov chain, from which posterior samples are drawn, multiple ``hot'' chains are maintained.
These hot chains typically move more freely through the parameter space.
The cold chain is periodically swapped with a hot chain to ``jump'' through the parameter space.

\subsection{Implementation}
We developed the open source software package sprspace (\url{https://github.com/cwhidden/sprspace}) to construct SPR graphs.
This software package also implements our clustering routine, prepares graph visualizations for Cytoscape, computes access times and commute times and computes our topological Gelman-Rubin-like measure.
Our software allows users to specify a fixed clustering radius in case dynamic cluster radius selection provides poor results.
Moreover, users may modify the number of top trees considered to change the amount of computation required.

\subsection{Data and run-time parameters}
\arxiv{\TABdatahohna}
We investigated MCMC estimation on unrooted trees by applying MrBayes 3.2 \citep{Ronquist2012-hi} to 17 empirical data sets.
The first group of data sets, which we will call DS1-DS11, have become standard data sets for evaluating MCMC methods \citep{lakner2008efficiency,hohna2012guided,larget2013estimation}.
These data sets consist of sequences from 27 to 71 eukaryote species (Table~\ref{TABdatahohna}), and are fully described elsewhere \citep{lakner2008efficiency}.
Note that TreeBASE identifiers for these data sets have changed from those used in some previous publications (Supplemental Table~\ref{TABlegacytreebase}).
The second group of data sets, which we will call VL1-VL6, consist of alignments with 40 to 63 bacterial and archaeal sequences (Table~\ref{TABdatahohna}) of protein-coding genes, and are fully described elsewhere \citep{beiko2006searching}.

To analyze the level of convergence to the posterior distribution, we computed large ``golden run'' posterior samples for each data set, meaning that we repeatedly ran the chains well past the typical number of iterations used for such analyses: for each of our 17 data sets, 10 single-chain MrBayes replicates were run for one billion iterations and sampled every 1000 iterations.
These replicates were not Metropolis-coupled.
We discarded the first 25\% of samples as ``burn-in'' for a total of 7.5 million posterior samples per data set, and assumed that this long burn-in period implied stationarity, i.e.\ that after burn-in the chain was sampling from the stationary distribution of the MCMC.
Following \citet{hohna2012guided}, we assumed these runs accurately estimated posterior split frequency distributions because of the extreme length of these Markov chains in comparison to our data size.
To test this assumption, we estimated the split frequency error between replicated golden runs (maximum standard error of any split) as in \citet{hohna2012guided} (see Table~\ref{TABdatahohna}).
The estimated split frequency error was below 0.06\% for each of our data sets, suggesting that the various golden runs are sampling the same split frequencies.
Moreover, commonly applied diagnostics implemented in the MrBayes \texttt{sumt} and \texttt{sump} tools satisfied common thresholds (Supplemental Table~\ref{TABgoldenconvergence}), including having a standard error of log likelihoods at most 2.11, maximum standard deviation of split frequencies at most 0.015 (0.007 for all but DS1), maximum Gelman-Rubin split PSRF values of 1.000, and the effective sample size (ESS; a measure of the number of samples correcting for MCMC autocorrelation) for the treelength parameter (the sum of branch lengths) exceeding 650,000.

We cannot similarly assume that these golden runs have accurately estimated the posterior probability of all topologies.
We do, however, assume that the golden runs have accurately estimated the posterior probability of high probability topologies, namely the top trees taken from the combined golden runs.
To test this assumption, we estimated the topological error between replicated golden runs (maximum standard error of the posterior probability of the top trees) for the eukaryote datasets, analogously to the split frequency error calculation (Supplemental Figure~\ref{FIGtopoerror} and Supplemental Table~\ref{TABtopoerror}).
The estimated standard error among high probability topologies was generally at least an order of magnitude smaller than the posterior probability, validating this assumption.
However, datasets DS9 and DS11 were notable exceptions as each topology was sampled exactly once, with no overlap between runs.
As such, we do not assume that the empirical distribution on topologies for DS9 and DS11 are close to their posterior distributions.

We ran MrBayes on each of our data sets with 10 replicates of a varying number of runs (2 through 8) and chains (1 or 4) until the runs had ASDSF less than 0.01 or a maximum of 100 million iterations.
We sampled these runs every 100 iterations and again discarded the first 25\% of samples.
We then compared the effect of these parameters on running time and error in practice, where error was measured by the root mean square deviation (RMSD) of split frequencies as compared to the golden runs.

\section{Results}

\subsection{The shape of tree posteriors and identification of peaks}

Distance SPR graphs of the combined golden run tree posteriors from the eukaryote alignments revealed a wide variety of posterior shapes (Supplemental Figure~\ref{FIGsprgraph}).
The shapes and complexity of these graphs were clearly not exclusively determined by the number of species or nucleotides in the data set.
Topological peaks were evident as large disconnected components (DS1, DS5, DS6) or sets of high probability trees separated by paths of low probability (DS4, DS7).
In particular, the trees with highest posterior probability in the two peaks of DS1 were separated by only two SPR operations but moving between these peaks required leaving the 95\% credible set.
Large subgraphs of lower probability trees appeared as interesting substructures (e.g. the ``tail'' on the right hand side of the DS8 graph).
No graph could be constructed for DS9 or DS11 as no topology was sampled twice and arbitrary 4096-node subsets were not adjacent in SPR space.

\arxiv{\FIGbeikograph}
Distance SPR graphs of the combined golden run tree posteriors from the ``VL'' bacterial and archaeal alignments also showed a wide variety of posterior shapes (Fig.~\ref{FIGbeikograph}).
Several posteriors were composed of clumps of trees with similar probability, as in data set DS7, which came from identical or near-identical sequences.
These also indicated small changes in uncertain areas of the trees that seldom affect their likelihood but drastically inflate the true 95\% credible set of topologies (Supplemental Table~\ref{TABmulti}).
We refer to this as the true credible set for brevity.
Dataset VL6 provided a striking example of peaks.
The 4096 most probable topologies (25.3\% credible set) formed 3 disconnected components and the 8192 most probable topologies (31.9\% credible set) showed only small paths of connectivity between the 3 peaks.
We focused on the eukaryote (``DS'') data sets in the remainder of our tests to focus our efforts, unless mentioned otherwise.

\arxiv{\FIGsprcluster}
Clustering regions of tree space by descending probability (see Methods) highlighted topological peaks and other interesting features (Fig.~\ref{FIGsprcluster}).
In addition to the peaky data sets (DS1, DS4, DS5, DS6, DS7) identified with unclustered graphs, DS10 appears to contain at least two peaks.
The disconnected sub-peaks of DS1 and DS6 contained the second cluster of both data sets and, thus, the most probable trees outside of the first cluster from each data set.
Conversely, the disconnected component of DS5 contained trees of relatively low probability.
In non-peaky data sets (e.g. DS3 and DS5) clusters expanded radially from the most probable tree, which indicates relatively easy MCMC mixing.

The number of unique topologies was greatly inflated by ambiguous relationships (Supplemental Table~\ref{TABmulti}).
For example, the posterior of data set DS7 had an interesting ``grid'' structure composed of clumps of 15 trees with similar topology and probability.
On closer inspection, trees within a clump differed only in the configuration of a subtree containing four nearly identical \emph{Microcebus rufus} sequences.
In fact, these sequences differed in four nucleotides, with one unique mutation per sequence, providing no distinguishing information and inflating the true credible set by a factor of 15 (the number of configurations of four taxa).
To verify this effect, we removed three of these four sequences, computed 10 new golden runs, and plotted the resulting tree space (Supplemental Figure~\ref{FIGdssevenunique}).
As expected we obtained the same structure, but with one tree per 15-node clump and proportional posterior probabilities.
The extreme flatness of DS9 and DS11 arose similarly.
The majority rule consensus tree for data set DS9 contained two 4-way multifurcations and one 5-way multifurcation.
Resolutions of these multifurcations occurred with approximately equal frequency, inflating the true credible set by a factor of $15 * 15 * 105 = 23,625$.
Much of the ambiguity was caused by a set of 4 identical sequences and a set of 3 identical sequences.
The remaining ambiguity seemed to arise from substantially similar sequences.
Similarly, the consensus tree for DS11 contained numerous multifurcations including a multifurcation with 12 edges.
The number of samples was insufficient to compare resolutions of this multifurcation and determine if each was equally likely.
However, 9 of the taxa involved had the same sequence, which alone inflated the true credible set by at least a factor of $2,027,025$, and this multifurcation likely inflated the credible set by orders of magnitude more.
Moreover, the posteriors of data sets DS5, DS6, and DS10 were also inflated by ambiguity.
In these cases, none of the sequences involved were identical and resolutions occurred with similar but not equal probability.

\arxiv{\TABdifficulty}
The shape of a posterior tree space explains the difficulty of sampling from that distribution (Table~\ref{TABdifficulty}).
Peaky distributions often required a large number of iterations to reach the ASDSF cutoff and/or had high error rates respective to other data sets with a similar number of taxa.
In particular, DS1 required the largest number of iterations to reach the ASDSF cutoff and had the second highest RMSD of split frequencies despite having the fewest number of species.
The number of credible trees and the radius of the tree space also appears to be a factor.
DS5 has a large, wide credible set and required a large number of iterations to reach the ASDSF cutoff.
DS7 has a smaller credible set and required relatively few iterations for the split frequencies to converge.
The high error rates of DS7, however, may indicate that the sub-peak or posterior shape caused the chain to stop prematurely.
Despite the large number of taxa and explored topologies of DS9 and DS11, these flat posteriors had low error rates and average times to achieve an ASDSF of 0.01.
To remove the effect of identical sequences, we ran 10 new MrBayes replicates of these two data sets with all but one member of each set of identical sequences removed (DS9-U and DS11-U).
Removing duplicate sequences reduced the number of iterations required to reach an ASDSF of 0.01 with little effect on error rates as compared to the DS9 and DS11 golden run splits with the corresponding taxa removed.

\subsection{Identifying bottlenecks in tree space}

\arxiv{\FIGdsonepeaks}
We were able to explicitly identify bottlenecks in tree space by examining SPR paths between high probability trees separated by regions of low probability.
As mentioned above, the most probable topologies of DS1's two peaks are separated by only two SPR operations.
However, these SPR operations have an inverted nested structure (Fig.~\ref{FIGdsonepeaks}).
The intermediate topology in this shortest path was so unlikely that it was never sampled in any of our tests.
This induces a severe bottleneck that results in the two peaks of DS1.
The peaks of DS6 arise from a different type of bottleneck (Supplemental Fig.~\ref{FIGdssixpeaks}).
Three SPR operations are required that move three subtrees into a common clade.
Both types of bottleneck are caused by a dependence between splits.

Topological peaks can lead to incorrect estimation of posterior distributions.
In addition to long times to achieve small ASDSF and high error rates, there is a risk of missing a peak entirely.
This was particularly evident for data set DS1 where 2 of our 10 tests with the MrBayes default settings failed to sample the sub-peak before reaching the ASDSF cutoff.
The cumulative posterior probability of this sub-peak (as calculated via golden runs) was approximately 20\%.
Four splits receive 95-99\% support when this sub-peak is missed as opposed to 75-80\% support (Supplemental Fig.~\ref{FIGdsoneconsensus}).

\subsection{Metropolis-coupling improves mixing between peaks}

\arxiv{\FIGweightedmcmc}
Metropolis-coupling~\citep{geyer1992practical,huelsenbeck2001mrbayes}, also known as MCMCMC, connected peaks together for these data sets (Fig.~\ref{FIGweightedmcmc}).
Weighted MCMC graphs of the peaky DS1 for posterior samples without Metropolis-coupling revealed that a single Markov chain rarely transitions between the peaks.
For example, there were only 4 observed transitions between peaks in one million-tree sample subsampled from an 100-million iteration MCMC run (Fig.~\ref{FIGweightedmcmc}(a)).
Given the large number of iterations and lack of Metropolis-coupling, it is unlikely that the chain frequently traversed between a peak and returned to the same peak between sampling periods.
MCMCMC, however, frequently jumps between the peaks.
In one approximately 1.2-million tree sample, subsampled from a 12-million iteration MCMCMC run with 4 chains (Fig.~\ref{FIGweightedmcmc}(b)), there were more than 4000 observed transitions between the central peak trees.
The effect of squashing these graphs together was more pronounced for the deep valley of DS1 as opposed to DS4 (Fig.~\ref{FIGweightedmcmc}(c)-(d)).

\arxiv{\FIGmct}
\arxiv{\TABcover}
To quantify mixing we computed \textit{commute time} statistics for each topology in the 95\% credible set; the commute time here was defined to be the number of Markov chain iterations necessary to move from the highest probability topology to the given tree and back.
The \textit{round trip cover time} is the number of iterations necessary to visit every topology in the credible set and return to the highest probability topology.
Metropolis-coupling also reduced the mean commute time (Fig.~\ref{FIGmct}) and round trip cover time (Table~\ref{TABcover}).
This effect was particularly pronounced for data set DS1.
The round trip cover time decreased by more than a factor of four for DS1, DS4, DS5, DS6, and DS8, outweighing the factor of four increase in computation time, whereas on data sets DS2, DS3, DS7, and DS10 the improved mixing rate of Metropolis-coupling did not outweigh the increased computation.
However, Metropolis-coupling reduced total computation time substantially, as the data sets where it did not reduce total computational effort to achieve a fixed ASDSF mixed relatively quickly compared to the ones for which it did.
Commute and cover time statistics could not be estimated for the flat DS9 and DS11 posteriors.
These results suggest that Metropolis-coupling does improve mixing between peaks and reduce total computational effort on average, but may not be beneficial for all posterior shapes.

Trees within sub-peaks were observed to have much larger commute times than other trees with a similar posterior probability.
This effect was particularly prominent in data sets DS1, DS4, DS6, and DS7 (Fig.~\ref{FIGmct}).
For example, the commute time of the central tree in the sub-peak of DS1 was 2.6 million iterations as opposed to 3,300-5,500 iterations for trees with similar probability.
This reduced to 80,200 and 2,100-3,700 iterations, respectively with Metropolis-coupling.
Similarly, the most probable trees within the three sub-peaks of DS4 (Fig.~\ref{FIGsprcluster}) had commute times between 200,000-307,000 iterations (37,000-47,000 with Metropolis-coupling).
Other trees of similar probability had commute times between 16,000-25,000 iterations.
Generally, our commute time analysis further demonstrates the difficulty of sampling sub-peaks and allows quantification of this difficulty.

\subsection{A small ASDSF when calculated with two runs is not always sufficient to ensure that empirical split frequencies are close to their posterior distribution; Metropolis-coupling aids in split frequency mixing}

As described in Methods, ASDSF compares split frequencies between runs (we emphasize that these runs are completely distinct and not coupled as for MCMCMC).
Increasing the number of simultaneous Markov chain runs greatly increased the stringency of a given ASDSF cutoff (Supplemental Figure~\ref{FIGerror}).
We found that ASDSF calculated using two runs is not sufficient for estimating the convergence of split frequencies.
Adding additional runs both increased the number of iterations required to reach the ASDSF cutoff and decreased the amount of error.
This effect varied by data set and peaky distributions saw the greatest decrease in error with additional runs.

In most cases, a small ASDSF implied that other convergence diagnostics were satisfied, regardless of the number of runs.
The mean potential scale reduction factor (PSRF; see Methods) for branch lengths was less than 1.01 in all but the 2-run DS2 and DS3 cases and 4 2-run DS7 cases, where the mean PSRF was less than 1.042.
Similarly, the ESS for the treelength parameter was greater than 200 except for data sets DS2 and DS3 and 8 of the 4-chain 2-run DS7 cases.

The ASDSF and split frequency error varied considerably over Markov chains of peaky data sets as runs transitioned between peaks.
These statistics often dipped below commonly applied thresholds only to increase rapidly after one run began exploring an alternative peak.
The subsequent rise and fall of these statistics decreased in magnitude as we gathered a sufficient sample.
However, current convergence diagnostics assume that these statistics decrease smoothly and, in particular, do not rise sharply.
The first time that an ASDSF cutoff is reached may not indicate that split frequency estimates are close to their posterior probabilities in peaky posteriors.
Moreover, the stopping time for Markov chains is often determined by the first occurrence of a sufficiently small split frequency deviation from golden runs \citep{hohna2012guided}.
This approach may underestimate the time needed to run these chains in the presence of topological peaks because the running observation of the split frequency may get close to the golden run split frequency because of stochasticity.

\arxiv{\TABdsoneconvergence}
Metropolis-coupling decreased error when peaky distributions were sampled with a small number of runs.
Dataset DS1, in particular, required 7 or more runs to achieve a mean RMSD below 0.02 without Metropolis-coupling but only 3 runs with Metropolis-coupling (Table~\ref{TABdsoneconvergence}).
Much of this error occurred when runs prematurely stopped using a split frequency criterion on the larger peak.
Even with 8 runs, one replicate without Metropolis-coupling reached an ASDSF of 0.01 after only 740,000 iterations, compared to the mean 82 million iterations.
None of the 8 runs visited any tree in the sub-peak, resulting in an RMSD of 0.08 and similar effects to those detailed above (Supplementary Fig.~\ref{FIGdsoneconsensus}).
The common diagnostics were satisfied for this replicate, including an ASDSF value less than 0.01, tree length ESS value of 3054, tree length PSRF of 1.000 and a maximum split frequency PSRF of 1.001.
Even with Metropolis-coupling, the MrBayes default of two runs was insufficient to adequately sample data sets DS1, DS4, and DS7 at the 0.01 ASDSF threshold.

\subsection{Topological Gelman-Rubin-like statistic}

\arxiv{\TABtopogr}
Because split frequency is a projection of the actual posterior on phylogenetic trees rather than the posterior itself, we wondered to what extent split-based measures being small implies that the empirical frequency on phylogenetic tree topologies is close to the posterior.
To explore this question, we developed a variant of the Gelman-Rubin statistic that used SPR distances (Methods).
This measure compares the mean square topology deviation within independent Markov chains to that between the chains.
The corresponding PSRF will approach 1 as the independent runs converge in topology distribution.

On our data sets, a small ASDSF generally implied that the topological measure was small (Table~\ref{TABtopogr}).
PSRF estimates with our topological Gelman-Rubin-like  measure approached 1, regardless of the number of runs.
Surprisingly, this also held for the flat posteriors of DS9 and DS11.
This suggests that similar trees were explored between runs of these posteriors, even if no two trees were identical.
There was little difference in topology deviation or PSRF with or without Metropolis-coupling.
Moreover, topological PSRF and ASDSF showed similar trends over time (Supplemental Fig.~\ref{FIGtopogr}), although the scale of this relationship appears to vary between different data sets and even different replicated tests on the same data set.

\subsection{Multidimensional scaling}

\arxiv{\FIGmds}
In general, MDS projections were insufficient to diagnose peaks (Fig.~\ref{FIGmds}) and extra information is required such as commute time, posterior density, and connectivity.
For flat posteriors, however, where extra information is unavailable, multidimensional scaling remains the only method of visualizing tree space (Supplemental Fig.~\ref{FIGmdsflat}).

Specifically, MDS plots often highlighted topological differences that did not impede mixing and missed sub-peaks.
For data set DS1, MDS displayed 4 clusters.
One axis separated the two peaks of DS1, but the other axis separated trees according to a common difference that did not impede mixing.
MDS plots of DS4 identified only one of three difficult to reach areas of tree space.
MDS plots were generally similar between RF and SPR.
In DS6, however, the plots differed significantly.
This highlighted the fact that the peaks of DS6 are quite close in SPR terms (despite their separation by a valley of improbable trees) but had very different splits.
We also compared MDS plots on the peaky microbial data set VL6.
Only one of the peaks was separated from the others in the SPR plot and the RF plot broke the peaks into multiple clumps.

\subsection{Effect of peaks on conditional clade probabilities}

\arxiv{\FIGccp}
Recent work uses a product of conditional posterior probabilities on splits as a proxy for the corresponding phylogenetic posterior probability \citep{hohna2012guided,larget2013estimation,szollosi2013efficient}.
This assumes an independence between the split probabilities of sister clades conditioned on their parent clade.
\citet{larget2013estimation} found several examples of trees where CCD probabilities differ from well-sampled empirical frequencies in the eukaryote datasets and where the simple estimates were well above the sampling threshold.
\citet{larget2013estimation} conducted two tests per eukaryote dataset using MrBayes 3.2 with the GTR model for 5,500,000 iterations, subsampling 100,000 trees with a 500,000-tree burn-in period.
He used these long runs as being representative of the posterior distribution, noting that ``A second set of runs under the same conditions, but with different random numbers, shows very similar results, indicating that these MCMC samples are likely not to suffer from poor convergence (data not shown).''
We extended his investigation of differences between CCD probabilities and well-sampled empirical frequencies with our substantially larger 1-billion iteration MCMC golden runs, subsampling 750,000 trees with a 250,000,000-tree burn-in period, replicated 10 times.

We found that conditional independence clearly did not hold in peaky distributions (Fig.~\ref{FIGccp}).
Specifically, conditional clade probabilities systematically underestimated the probability of trees within sub-peaks and overestimated the probability of trees between the peaks.
This effect was exemplified in data set DS1 where highly unlikely trees between peaks had conditional clade probabilities exceeding one percentage point (points significantly above the line in Figure~\ref{FIGccp}(a)).
We observed similar effects in DS4 and DS6.
Surprisingly, even in the relatively simple posterior of DS3, CCD underestimated the posterior probability of three trees in the 95\% credible set by an order of magnitude.
However, CCD performed well overall on non-peaky data sets and is currently the only way to estimate probabilities below the sampling threshold (e.g. DS9 and DS11).

\section{Discussion}

We developed the first practical method for examining the subtree prune-and-regraft tree space of Bayesian phylogenetic posteriors.
Our novel graph-based approach uses size and color to visualize connectivity, posterior probability, and relative distance.
Our simple clustering procedure identified topological peaks in several real data sets.
Additionally, we investigated the impact of Metropolis-coupling, the number of runs used for ASDSF calculation, and developed a convergence diagnostic that uses phylogenetic tree topologies directly.

We find that multimodal or ``peaky'' posteriors are common in data sets with 30 or more taxa, confirming the suggestion by \citet{beiko2006searching}.
Markov chains on peaky posteriors often required a large number of iterations to obtain small ASDSF values and had high error rates relative to the number of taxa.
We used dynamic programming to compare tree commute times and found that trees within sub-peaks were difficult to sample.
The ``height'' of a peak compared to the ``depth'' of the corresponding valley influenced sampling difficulty.
Dataset DS1, despite its relatively small number of taxa, has a large sub-peak separated by a particularly deep valley.
In many cases, this led to premature termination of chains by the ASDSF measure and erroneously assigning greater than 95\% confidence to some relationships with an actual frequency less than 80\%.

We explicitly identified tree space bottlenecks in two data sets with tall sub-peaks and found that they were caused by a dependence between splits.
These peaks were only isolated by a handful of SPR operations.
However, the intermediate valley topologies were exceedingly unlikely and the SPR operations modified a large number of splits.

Dependence between sister clades caused systematic errors in CCD probability estimates.
Specifically, CCD overestimated the probability of trees between peaks and underestimated the probability of trees in sub-peaks.
These observations suggest that CCD-guided proposal operators could hide sub-peaks and further aggravate the difficulty of sampling peaky phylogenetic posteriors.
On the other hand, CCD-guided proposal operators may sample valley trees more frequently and therefore provide more chances to cross valleys and sample sub-peaks.
Tree space sampling methods that penalize or even prevent SPR and TBR operators that change a large number of splits could also hide sub-peaks, such as the ``pruning distance'' of \citet{hohna2012guided} and similar suggestions \citep{huelsenbeck2008bayesian,lakner2008efficiency}.
Moreover, an anti-peak bias would be undetectable and, perversely, decrease running times using an ASDSF rule or other common convergence diagnostic.
One strategy to alleviate bias, while still retaining the benefit of CCD, might be to use CCD or other biased proposal distributions in a subset of Markov chains (Metropolis-coupled or otherwise) along with chains using a general proposal distribution.
CCD has also begun to see use in phylogenomic methods such as amalgamated likelihood estimation \citep{szollosi2013efficient}, which uses CCD directly as a proxy for posterior probability in order to infer a species tree joint with a set of gene trees.
Biases in CCD will bias the results of this approach.

Identical and closely related sequences cause ambiguity which greatly inflated the tree space sampled by MCMC methods.
Such data sets had large and flat posteriors, which were difficult to quantify and visualize.
Ignoring duplicate sequences reduced mean run time (determined by the ASDSF stopping rule) by 26\% and 50\% in our two flattest posteriors.
Thus, we suggest that users of MCMC should identify and ignore duplicate sequences, maintaining only a single representative from each set of identical sequences.
The post-processing of such an analysis could either expand the representative to a multifurcating clade containing each ignored sequence from a set, or spread the probability of each sampled tree uniformly across each full tree with monophyletic clades for the expanded sets.
Ignoring duplicate sequences may make branch length priors harder to interpret, due to a consequent ascertainment bias.
This may make little difference in practice, however, as commonly used priors do not allow for large numbers of simultaneous or nearly simultaneous branching events.
Intelligently handling duplicate sequences may be a useful feature of future MCMC software.
Moreover, future work should explore methods for handling closely related sequences without inflating tree space.
This could be done with reversible jump Markov chain Monte Carlo \citep{lewis2005polytomies}.
Tree space inflation will be of particular importance when estimating trees for a large number of closely related sequences as in personalized medicine and metagenomics.

Metropolis-coupling was effective in reducing commute times and decreased the mean cover time of the 95\% credible set by more than a factor of 4 in peaky distributions.
Metropolis-coupling increased the number of transitions between peaks by three orders of magnitude in our peakiest data set.
However, Metropolis-coupling may not be effective for all posterior shapes.
The observed cover time decrease in non-peaky data sets did not outweigh the increased computation of Metropolis-coupling, however because Metropolis-coupling significantly reduced computational load for the most difficult and time-consuming posteriors, it appears to be a useful default option on average.
We tested the effect of Metropolis-coupling with 4 chains, the default number of chains in MrBayes, and future work should investigate the optimal number of Metropolis-coupled chains for peaky and non-peaky posteriors.
Moreover, further research could investigate whether it is heating, multiple chains, or both that improves mixing in peaky posteriors.

The magnitude of the ASDSF convergence diagnostic depends heavily on the number of Markov chains used for comparison.
We found that using 2 independent runs with an ASDSF cutoff of 0.01 resulted in insufficient chain lengths for peaky posterior distributions.
Indeed, MCMC runs often stopped using a 2-run ASDSF stopping rule before sampling a sub-peak.
This is a serious consideration, as MrBayes uses 2 runs by default, and MrBayes uses a default ASDSF termination threshold of 0.05 when ASDSF termination is enabled but no threshold provided.
Moreover, MrBayes does not provide a warning message unless the ASDSF exceeds 0.1 when run for a fixed number of iterations.
ExaBayes uses ASDSF termination by default with a threshold of 0.05.
We did not test similar single-chain convergence diagnostics (e.g the methods of \citet{raftery1992many} or \citet{geweke1991evaluating}) but they may experience similar problems.
MCMC analyses should use at least 3 independent runs and an ASDSF threshold of at least 0.01 in any MCMC analysis for which accurate topological posterior estimation is an important concern.
Moreover, multiple independent MCMC replicates should be compared---using even 8 runs was not enough to prevent one of our MCMC tests from stopping with an ASDSF stopping rule before sampling the sub-peak of DS1.
A wide variance in chain lengths using split frequency stopping rules on independent replicates may be a sign of topological sub-peaks.

We developed a topological Gelman-Rubin-like convergence diagnostic which works directly on tree topologies.
This diagnostic can be applied with any distance metric on tree topologies.
Tests with this topological Gelman-Rubin-like measure suggest that small ASDSF often implies a small topological Gelman-Rubin-like diagnostic for high-probability topologies, although neither measure can detect unsampled topological peaks.

A major and natural difficulty of peak detection is that the peaks must be sampled in order to be detected.
Similarly, it is difficult to accurately estimate time to satisfy some convergence criterion.
Convergence time estimates using golden runs \citep{hohna2012guided} are based on the first time that split frequencies of a Markov chain approach the golden run split distribution.
However, this approach may underestimate the running time of MCMC methods in practice because sampled split frequencies can approximately hit the golden run split distribution before they have stabilized.
It may be worth checking that split distributions have stabilized in addition to requiring them to hit the golden run split distribution.

Our methods could be expanded in several ways.
We limited many of our comparisons to subsets of at most 4096 trees due to the computational overhead of pairwise comparisons.
Our approach would benefit greatly from faster methods for unrooted SPR comparisons or a way to construct SPR graphs without comparing every pair of trees.
There also are special challenges in moving through the space of rooted trees with a time component (as estimated by BEAST), which would be interesting to investigate; our methods would also be much more efficient on posteriors of these rooted trees.
We developed a very simple method for highlighting topological peaks that was designed to dynamically select cluster radii with few SPR comparisons.
Our clustering procedure worked well in our tests, but in multiple situations could select unreasonably small cluster sizes (e.g. if the standard deviation approached or exceeded the mean).
Improved methods for identifying such peaks and analyzing tree space graphs should be explored.
In particular, methods are needed to rapidly scan posteriors for common bottlenecks in order to develop new phylogenetic operators that cross those bottlenecks.
Moreover, future work should determine the cause of such bottlenecks in terms of sequence features (for example mixtures of tree topologies).
Nontrivial methods will be required to do so in the likelihood-based framework.
Finally, our observations need to be confirmed on other data sets.
This work is but a first step in quantifying MCMC exploration of phylogenetic tree space using topological methods.

\section{Funding}
This work was supported by the National Science Foundation (grant number 1223057).

\section{Acknowledgements}
We would like to thank Connor McCoy for some initial simulation work and helpful discussion, as well as the helpful comments and discussion of Aaron Darling.
We would also like to thank Robert Beiko for his comments and data set suggestions, as well as the phylobabble.org community for helpful discussion.
Editor-in-Chief Frank (Andy) Anderson, Associate Editor Mark Holder, Bret Larget, Tracy Heath and an anonymous reviewer all provided helpful suggestions as part of the review process.

\bibliographystyle{sysbio}
\bibliography{sprmix}

\clearpage
\notarxiv{
\section{Tables}

\TABdatahohna
\TABdifficulty
\TABcover
\TABdsoneconvergence
\TABtopogr

\setcounter{table}{0}

\renewcommand{\tablename}{Supplemental Table}
\TABlegacytreebase
\TABgoldenconvergence
\TABtopoerror
\TABmulti

\clearpage

\section{Figure captions}

\clearpage

\FIGspr
\FIGbeikograph
\FIGsprcluster
\FIGdsonepeaks
\FIGweightedmcmc
\FIGmct
\FIGmds
\FIGccp

\setcounter{figure}{0}
\renewcommand{\figurename}{Supplemental Figure}

\FIGusprcounterexample
\FIGtopoerror
\FIGsprgraph
\FIGdssevenunique
\FIGdssixpeaks
\FIGdsoneconsensus
\FIGtopogr
\FIGerror
\FIGmdsflat

}

\arxiv{
	\section{Supplementary Material}
	\setcounter{figure}{0}
	\setcounter{table}{0}
	\ifdefined\notarxiv
\newcommand{\included}[1]{}

\else

\newcommand{\included}[1]{#1}

\documentclass[12pt,letterpaper]{article}

\usepackage[round]{natbib}
\usepackage{graphicx}
\usepackage{url}
\usepackage{enumerate}
\usepackage{bm}
\usepackage{color}
\usepackage{palatino}
\usepackage{csvsimple}
\usepackage{subfigure}
\usepackage{fullpage}
\usepackage{amsmath,amsfonts,amssymb,amsthm}
\usepackage[labelfont=bf,labelsep=period,font=small]{caption} 


\newcommand{\PP}{\mathbb{P}}
\newcommand{\EE}{\mathbb{E}}
\newcommand{\RR}{\mathbb{R}}
\newcommand{\ZZ}{\mathbb{Z}}
\newcommand{\Multinom}{\operatorname{Multinom}}
\newcommand{\Dirichlet}{\operatorname{Dirichlet}}


\newcommand{\arxiv}[1]{#1}
\newcommand{\notarxiv}[1]{}
\newcommand{\eat}[1]{}

\newtheorem{lemma}{Lemma}
\newtheorem{prop}{Proposition}
\newtheorem{theorem}{Theorem}
\newtheorem{problem}{Problem}
\newtheorem{defn}{Definition}
\newtheorem{obs}{Observation}
\newtheorem{alg}{Algorithm}

\newcommand{\EM}[1]{\textit{\colorbox{magenta}{Erick:} #1}}
\newcommand{\CW}[1]{\textit{\textcolor{white}{\colorbox{blue}{Chris:}} #1}}

\input{figures.tex}

\title{Quantifying MCMC exploration of phylogenetic tree space -- Supplementary Material}
\author{Christopher Whidden \and Frederick A. Matsen IV}
\date{\today}

\renewcommand{\section}[1]{%
\bigskip
\begin{center}
\begin{Large}
\normalfont\scshape #1
\medskip
\end{Large}
\end{center}}
\renewcommand{\subsection}[1]{%
\bigskip
\begin{center}
\begin{large}
\normalfont\itshape #1
\end{large}
\end{center}}
\renewcommand{\subsubsection}[1]{%
\vspace{2ex}
\noindent
\textit{#1.}---}
\renewcommand{\tableofcontents}{}
\bibpunct{(}{)}{;}{a}{}{,}  
\linespread{1.6}
\raggedright
\setlength{\parindent}{0.5in}
\renewcommand\bibsection{\section{\refname}}

\begin{document}

\maketitle

\fi

\renewcommand{\tablename}{Supplemental table}
\renewcommand{\figurename}{Supplemental figure}

\section{Computing access time statistics}
\label{sec:computing_MAT}

The MAT between two trees $i$ and $j$, $\operatorname{MAT}(i,j)$, is the mean number of iterations before node $j$ is visited after node $i$ is visited.
The mean commute time $\operatorname{MCT}(i,j)$ is the mean number of iterations for a random walk to visit tree $i$, then tree $j$, and finally return to tree $i$.
The MCT for two trees can be computed as $\operatorname{MCT}(i,j) = \operatorname{MAT}(i,j) + \operatorname{MAT}(j,i)$.
The $\operatorname{MRT}(i, T)$ is the mean number of iterations required to cover (visit) each tree in set $T$ starting from tree $i$ and then return to tree $i$.
The MRT of a graph $T$ is the maximum of $\operatorname{MRT}(i, T)$ across nodes $i$.
The MRT can be computed as the maximum MCT from node $i$ to a tree $t$ in $T$.

The MAT values (and hence MCT and MRT values) involving the highest probability tree, $t_0$, can be computed with a single pass through the tree posterior using a method for updating weighted means.
To do so, we use dynamic programming and store three values: (1) $c_{ij}$, the number of times a topology $j$ has been seen since the last visit to topology $i$, (2) $m_{ij}$, the mean iteration number of each such visit, and (3) the current $\operatorname{MAT}(i,j)$ estimate.
We perform updates when one of $i$ and $j$ is $t_0$ as follows.
For each posterior sample $j$ with $j = t_0$, we update our values for each topology $i$.
We update the access time estimates $\operatorname{MAT}(i,j)$ with weight $c_{ij}$ and value $m_{ij}$ and then reset the weight and value.
We then update the stored values $c_{ji}$ and $m_{ji}$.
If $j \ne t_0$, we apply the same update procedure but only for $i = t_0$.
This requires linear storage with respect to the number of distinct compared topologies.

\pagebreak

\TABlegacytreebase
\TABgoldenconvergence
\TABtopoerror
\TABmulti
\FIGusprcounterexample
\FIGtopoerror
\FIGsprgraph
\FIGdssevenunique
\FIGdssixpeaks
\FIGdsoneconsensus
\FIGerror
\FIGmdsflat
\FIGtopogr

\included{
	
}

\end{document}